\documentclass[12pt,a4paper]{article}
\usepackage{graphicx}
\usepackage{booktabs}
\usepackage{dcolumn}
\usepackage{natbib}
\usepackage{xcolor}
\usepackage{bm}
\usepackage[left=2.6cm,right=2.5cm,top=2.5cm]{geometry}

\bibpunct[: ]{(}{)}{;}{a}{}{,}

\begin{document}

\title{A Theoretical and Empirical Comparison of the Temporal Exponential Random Graph Model and the Stochastic Actor-Oriented Model\footnote{The authors wish to thank the many people who have provided extensive comments on this paper. All errors remain our own. \medskip \newline PL gratefully acknowledges support from the Zukunftskolleg at the University of Konstanz. SJC gratefully acknowledges support from the National Science Foundation (SES-1514750, SES-1461493, and SES-1357622) and the Alexander von Humboldt Foundation. \medskip \newline Address for correspondence: Philip Leifeld, University of Glasgow, School of Social and Political Sciences, Adam Smith Building, Bute Gardens, Glasgow, G12 8QQ, Scotland, United Kingdom. E-mail: \texttt{philip.leifeld@glasgow.ac.uk}.}}
\author{Philip Leifeld\\\small{University of Glasgow} \and Skyler J.\ Cranmer\\\small{The Ohio State University}}
\date{}

\maketitle

\begin{abstract}
The temporal exponential random graph model (TERGM) and the stochastic actor-oriented model (SAOM, e.\,g., \texttt{SIENA}) are popular models for longitudinal network analysis. We compare these models theoretically, via simulation, and through a real-data example in order to assess their relative strengths and weaknesses.  Though we do not aim to make a general claim about either being superior to the other across all specifications, we highlight several theoretical differences the analyst might consider and find that with some specifications, the two models behave very similarly, while each model out-predicts the other one the more the specific assumptions of the respective model are met.
\end{abstract}

\section{Introduction}
When modeling longitudinally observed networks (e.\,g., panel network data), the analyst is generally confronted with a choice between two popular candidate models: the temporal exponential random graph model (TERGM) \citep{Hanneke:2010, Leifeld:2018, Desmarais:2010nips, Desmarais:2012physa}
and the stochastic actor-oriented model (SAOM) \citep{Snijders:2001, Snijders:2007, snijders2010introduction}, often known by the name of its software implementation \texttt{SIENA}. An unusual, even surprising, feature of the literature on longitudinal network analysis is that authors tend to use one or the other of these techniques without explicitly considering or refuting the alternative. This is noteworthy as previous studies show that the two models often yield considerably different parameter estimates based on nearly identical model specifications \citep{lerner2013conditional}. Even though the two models accomplish the same inferential task in a somewhat similar way, it is rare to see the theoretical fit or empirical performance of the two techniques directly compared \citep[but see][]{block2018change, block2018forms}. When theory is invoked with respect to model choice, it is usually done to justify the use of one of these network models relative to a regression model.\footnote{There is a substantial literature on why regression models are inappropriate for network data. That is not a subject we will examine here as it has already been well covered elsewhere. This literature begins with reviews of the work by Sir Francis Galton \citep{dow1984galton}, but also includes more recent work, such as \cite{hoff2002latent}, \cite{hoff2004modeling}, \cite{Goodreau:2008}, \citet{lerner2013conditional}, and
\cite{Cranmer:2011}, and \citet{cranmer2017navigating}. }
Yet, as we demonstrate below, it is usually not the case that theory alone is strong enough to provide the analyst a clear preference for one or the other. So, what is one to do?

Our goals here are to (a) carefully discuss the differences between the TERGM and SAOM such that the reader may be able to consider the interface between the two models and substantive theory in any given application area. We also (b) use a simulation study to show that differences in the models' assumptions and mechanics often result in significantly different substantive conclusions, even when the same theoretical specification is tested on the same data. We simulate two different types of known data-generating processes catering to the different assumptions each model makes. The results show that a violation of these assumptions generally leads to a larger discrepancy between actual and predicted edges and that the models are sensitive to misapplication due to their different temporal updating processes. Further (c), we seek to demonstrate the utility of comparing the two competing models in application, as it is usually unclear \emph{a priori} which of the two models' assumptions are met more exactly. To this end, we replicate a popular SAOM specification with a TERGM and compare the out-of-sample predictive performance of the models. The comparison illustrates that even in well-studied applications, the appropriateness of the respective model-related assumptions can be hard to determine theoretically. Lastly (d), we provide a framework and companion software for easily comparing the out-of-sample predictive performance of the TERGM and SAOM. Thus, researchers whose theory does not provide sufficient guidance to select a model, or researchers seeking to evaluate the robustness of their results, need not rely on theory alone to judge which model fits best for a given application.

\section{Theoretical Comparison Between the Two Models}

\subsection{Common Roots of the SAOM and the TERGM}
The inferential study of temporal dynamics in networks goes back to the 1970s and 1980s, with pioneering work by \citet{holland1977dynamic}, \citet{iacobucci1988general}, \citet{wasserman1996logit}, \citet{robins2001random}, \citet{runger1980longitudinal}, and \citet{wasserman1986statistical, wasserman1988sequential}. Both models we consider here were inspired by this literature.\footnote{Both models are also related to the relational event model (REM) \citep{butts2008relational} and the dynamic network actor model (DyNAM) \citep{stadtfeld2017interactions}, two separate longitudinal network models that are applicable to temporally fine-grained sequences of edges (for a comparison of the two, see the debate by \citealt{stadtfeld2017dynamic}, \citealt{butts2017comment}, and \citealt{stadtfeld2017rejoinder}). While the four models share the property that one can incorporate sufficient statistics for capturing network dependence, the REM and DyNAM are not considered further in our comparison because they are not applicable to panel network data.}

A natural starting point in comparing the TERGM and the SAOM is an account of their many similarities, followed by a comparison of their differences. The mathematical hearts of the TERGM and SAOM are very similar and are both related to the (non-temporal) exponential random graph model (ERGM) \citep{wasserman1996logit}. The models differ only in their sets of possible permutations, their treatment of temporal dynamics, and their estimation strategies (discussed in the next two sections).

The (non-temporal) ERGM can be expressed by its probability density function
\begin{equation}
 P(N, \bm{\theta}) = \frac{1}{\sum_{N^* \in \mathcal{N}} \exp\{\bm{\theta}^\top \textbf{h}(N^*) \} } \exp\{\bm{\theta}^\top \textbf{h}(N) \},
\label{ergm}
\end{equation}
where $N$ is the observed network, $\bm{\theta}$ are the parameters, $\textbf{h}(N)$ is a vector of statistics computed on the network, and $N^*$ refers to a particular permutation of the network from the set of all possible permutations of the network holding the number of vertices as fixed $\mathcal{N}$. Here, a permutation is defined as another network topology with the same number of vertices but not necessarily the same number of edges. Equation~\ref{ergm} yields the exponentiated sum of weighted network statistics over the same sum for all networks that could have been observed in place of the network being considered in the numerator. A wide variety of endogenous dependence can be incorporated into the $\textbf{h}(N)$ vector of statistics, including reciprocity, edge-wise shared partners, four-cycles, and exogenous covariates, among many others \citep{Morris:2008}. The TERGM and the SAOM are both closely related to this definition, as will be explored below. For reviews of the ERGM, its specification, and its estimation, see, for example, \citet{Goodreau:2008},
\citet{Cranmer:2011},
and \citet{Lusher:2013}.

\subsection{The Temporal Exponential Random Graph Model}
The TERGM builds on the ERGM by first defining the probability of a network at the current time step $t$ as a function of not just sums of subgraph counts of the current network, but also previous networks up to some time step $t-K$:
\begin{equation} \label{tergm_t}
 P(N^{t}| N^{t-K}, \ldots, N^{t-1}, \bm{\theta}) = \frac{\exp(\bm{\theta}^\top \textbf{h}(N^{t}, N^{t-1}, \ldots, N^{t-K}))}{c(\bm{\theta}, N^{t-K}, \ldots, N^{t-1})}.
\end{equation}
This assumes that the statistics formed based on the networks between $t-K$ and $t$ fully encompass the dependencies observed in the network at time $t$. The denominator contains the same kind of normalizing constant as in the ERGM.
In a second step, one computes the product over all time periods in order to determine the probability of the time series of networks:
\begin{equation}
 P(N^{K+1},\ldots, N^{T} | N^{1},\ldots,N^{K}, \bm{\theta}) = \prod_{t=K+1}^{T} P(N^{t}|N^{t-K}, \ldots, N^{t-1}, \bm{\theta}).
\end{equation}
This is a simple extension of the ERGM to a series of networks. To incorporate temporal dependencies between time steps, the network statistics defined in $\textbf{h}$ can incorporate ``memory terms.''
\citet{Leifeld:2018}
provide an extensive discussion of this topic. The key is to specify a temporal statistic related endogenously to the structure of the network that captures the temporal process at work. A very simple example would be the dyadic stability term,
\begin{equation}
 \textrm{h}_m = \sum_{i \neq j} N_{ij}^t N_{ij}^{t-1} + (1 - N_{ij}^t)(1- N_{ij}^{t-1}),
\end{equation}
which adds value to the $\textrm{h}_m$ statistic whenever the status of a dyad (tied or not) does not change from one period to the next. Because the temporal processes of interest will vary from application to application, it is fortunate that the analyst has a great degree of flexibility in the design of memory terms. The only real restriction is that they should be specified as sums of subgraph products. As such, including terms to capture edge persistence, edge innovation, edge dissolution, delayed reciprocity, delayed triadic or four-path closure, delayed two-paths, or arbitrary functions of time, is possible
(\citealt{Hanneke:2010}; 
\citep{Hanneke:2010, Leifeld:2018}.

It is important to note that these dependencies comprise both autoregressive specifications, where the current network is modeled as a function of previous states of the network (such as dyadic stability) and network statistics that describe the transition between consecutive network states (such as delayed reciprocity or closure of transitive triads over two or three time steps), where the latter class of statistics permits modeling network evolution as a temporal process. In other words, the dependence graph \citep{frank1986markov} of the TERGM, which formally defines the dependence of one dyad on another, can (but need not) model the dependence between modelled variables at multiple different time points.

This makes the TERGM a flexible model because there are no restrictions on the dependencies other than the length of the time series. The TERGM does not make any assumptions as to whether the time that passes between time steps is long or short, continuous or discrete, and whether edges can be established sequentially or simultaneously in the data-generating process, \emph{as long as the outcome of this process can be mapped into a dependency term} that can be incorporated into the $\textbf{h}$ vector. This makes it very flexible because there are few restrictions on the $\textrm{h}$ statistics. Conversely, it does not maintain a very close connection to the data-generating process at the micro-level, as will be discussed below.

Estimated parameters can be interpreted as the log-odds of establishing a tie (at any point) given the rest of the network and up to $K$ previous networks. Estimation is possible using Markov chain Monte Carlo maximum likelihood estimation (MCMC-MLE) (\citealt{Hanneke:2010}; 
\citep{Hanneke:2010}
or maximum pseudolikelihood estimation with bootstrapped confidence intervals
\citep{Desmarais:2010nips, Desmarais:2012physa},
as implemented in
the \texttt{btergm} package \citep{leifeld2017xergm:, Leifeld:2018} for the statistical computing environment \texttt{R} \citep{coreteam2016r:}.
For a more detailed methodological treatment of the TERGM, see \citet{Hanneke:2010}, 
\citet{Leifeld:2018}, \citet{Cranmer:2011}, \cite{Desmarais:2012physa}, \cite{Cranmer:2012}, and \cite{Cranmer:2014}.

\subsection{The Stochastic Actor-Oriented Model} \label{saom}
To build a SAOM \citep{Snijders:2001}, one takes the initial observed network as given and models changes to it using a variant of Equation~\ref{ergm} and a simulation process meant to mimic the evolution of the network between discretely observed time periods. The SAOM is a stochastic process in continuous time and can be characterized as a generative model that describes the network dynamics over time. While the TERGM is primarily a joint model of network states at multiple time points and is usually parameterized in a way that builds dynamics across time points into the model as an explanatory factor, the SAOM models the changes that take place between time points, rather than the outcomes of these processes as in the TERGM. This, coupled with its explicit micro-level orientation, makes the SAOM a theoretically appealing choice for model building on network dynamics.

Ultimately, both techniques try to model a series of consecutive networks using endogenous network statistics and a set of covariates. In fact, the TERGM and SAOM include endogenous dependencies and exogenous covariates as subgraph products in the $\textbf{h}(N)$ term, in an almost identical fashion. Thus, the TERGM and SAOM can be characterized as very similar models. As will be made clear presently, the major difference between them stems from the ground assumptions of the models and the resultant differences in updating processes employed by the two models, as well as the support of their probability distributions.

The SAOM follows \citet{holland1977dynamic} in positing that network change is a first-order Markov process, meaning that the network at time $t$ is exclusively a function of the network at $t - 1$, a vector of endogenous statistics ($\textbf{h}$, which can be nearly identical to the $\textbf{h}$ vector in the TERGM), and a corresponding set of parameters $\bm{\theta}$ that determine the evolution of the network between $t-1$ and $t$. The SAOM posits that the time between $t-1$ and $t$ can be broken down into a possibly infinite number of consecutive time steps---so-called mini-steps--- during which vertices (often interpreted as actors) re-allocate their edges. At each of these mini-steps, two functions are executed: the rate-of-change function and the objective function.

At each mini-step, the \emph{rate-of-change function} first selects an actor that may change its tie structure. The rate of change depends on the network and individual characteristics. By default in the software and most commonly in the literature, the rate is specified as:
\begin{equation} \label{rate_function}
 \forall i:\quad\lambda_i(N^t) = \rho_t.
\end{equation}
This rate, or waiting time, is a random Poisson process that assigns an equal probability of being selected to each vertex at a given mini-step. That is, each vertex has the same expected waiting time before it is selected (again). The rate function can be specified in more elaborate ways such that vertices are activated with individual waiting times corresponding to functions of vertex covariates. When using a weighted rate-of-change function, characteristics affect the rate-of-change function multiplicatively through an exponential link function \citep{snijders2010introduction}. This choice should be driven by theory; the simplest option is an equal probability with waiting times scaled to match the amount of change between the observed networks at $t-1$ and $t$ \citep{Snijders:2001}. Details on the rate-of-change function can be found in \citet{Snijders:2001, snijders2005models}, \citet{amati2011new}, and \citet{snijders1997simulation}.

Given that an actor has been selected at a mini-step, the \emph{objective function} is executed. The actor selected may add a new edge to another vertex, remove an existing edge, or leave its edge profile as it is. The choice of which outgoing dyad $k$ to consider (from the perspective of vertex $i$) is guided by the function
\begin{equation} \label{objective}
f_i(\bm{\theta}, N) = \sum_k \theta_k \textrm{h}_{ik}(N).
\end{equation}
This function accommodates the same $\textbf{h}$ statistics as the ERGM, except that SAOM statistics are computed from an egocentric perspective. For example, while the (T)ERGM would include reciprocity as the count over all $ij$ dyads,
\begin{equation} \label{recip_ergm}
 \textrm{h}_{\textrm{reciprocity\_ergm}} = \sum_{i \neq j} N_{ij} N_{ji},
\end{equation}
the SAOM would include reciprocity from the perspective of vertex $i$,
\begin{equation} \label{recip_saom}
 \textrm{h}_{\textrm{reciprocity\_saom}} (i) = \sum_{j} N_{ij} N_{ji}.
\end{equation}
Apart from this difference in the formulation of statistics, any network statistic that can be included in the ERGM can also be included in the SAOM.

The decision of which dyad to change, as given in Equation~\ref{objective}, is made simultaneously with the decision on how to change the respective dyad. Intuitively, given that the objective function concerns an $n$-category unordered choice (consisting of the three types of actions: add or remove an edge; do nothing), the probability of a specific choice is computed in an identical fashion to that of a multinomial logistic regression:
\begin{equation} \label{changeprob}
 Pr(N_{ij}) = \frac{\exp (f_i(\bm{\theta}, N)) }{\sum_{N^* \in \mathcal{N}} \exp (f_i (\bm{\theta}, N^*))}.
\end{equation}
Put more simply, ``the probability that an actor makes a specific change is proportional to the exponential transformation of the objective function of the new network, that would be obtained as the consequence of making this change'' \citep[p.~58]{snijders2010introduction}. This may look identical to the probability definition of the ERGM in Equation~\ref{ergm}, but there are two notable differences: $\mathcal{N}$ contains only the network states that are under direct influence of vertex $i$ through the objective function (as opposed to all permutations in the ERGM), and the dependency terms $\textrm{h}_{ki}(N)$ in $f_i(\theta, N)$ are formulated from the perspective of $i$ (see Equations~\ref{recip_ergm} and~\ref{recip_saom}).

Estimated parameters in the SAOM denote contributions to the objective function, and log odds ratios can be retrieved by comparing the objective function for different outcomes (establish a tie, remove a tie, or do nothing) \citep{snijders2005models, snijders2010introduction}. The SAOM can be estimated using the (generalized) method of moments by means of a (modified) Robbins-Monro stochastic approximation algorithm \citep{Snijders:2001, amati2011new}. The model was implemented in the \texttt{R} package \texttt{RSiena} \citep{ripley2017rsiena, ripley2017manual}.

As this discussion has shown, the SAOM and TERGM are similar in many respects related to their core equations and specifications. Under a specific and rarely applied rate function, the SAOM even has an ERGM as its limiting distribution \citep{Snijders:2001}. A comparison of Equations~\ref{tergm_t} and~\ref{changeprob} with Equation~\ref{ergm} shows how both have strikingly similar mathematical cores with a nearly identical way of specifying sufficient network statistics.

\subsection{Differences between the TERGM and the SAOM} \label{sect:diffs}
The similarity between these models may lead to some confusion as to which model is preferable for a given application. In line with the ``no-free-lunch theorem'' \citep{boulesteix2013plea}, we believe that one cannot make a general claim about one of these models being generally superior to the other across contexts, but that the question of which model to choose must be addressed with respect to the specific application. Consequently, we review the assumptions of each model before considering the effects that these differences have on model performance.

Before we do so, however, we note that the SAOM and the TERGM each have their own recent or ongoing extensions, which may offer advantages over each other in specific contexts. The SAOM, for example, is able to estimate multiple interdependent network processes in a single model \citep{snijders2013model}. There are variants for ``multiplex'', ``multi-relational'', or ``multi-level'' ERGMs as well \citep{wang2013exponential}, but they have not yet been extended to the temporal case. The ERGM has also seen the development of variants for weighted (valued) edges 
(e.\,g., the Generalized ERGM) \citep{desmarais2012statistical},
and extensions to the temporal case should be straightforward. Here, we focus on the simplest case: a comparison between the SAOM and the TERGM for a single, binary network over multiple time periods as described by
\citet{Hanneke:2010}, \citet{Leifeld:2018}, \citet{Desmarais:2010nips}, and \citet{Desmarais:2012physa} and implemented in \citet{leifeld2017xergm:}
for the TERGM and as described in \citet{Snijders:2001}, \citet{Snijders:2007}, and \citet{snijders2010introduction} and implemented in \citet{ripley2017manual} for the SAOM. We consider these the canonical forms of the two models and limit our discussion to them. Contrasting the various extensions of these models is left to future research.

\subsubsection*{Primacy of Actors and Micro-Level Modeling}
One may be tempted, based on the difference in names, to think that the major theoretical difference between the SAOM and TERGM relates to the primacy of actors (the vertices of the network). In fact, the major difference lies not so much in the primacy of actors, but more in the timing and updating processes assumed in the two models. However, it is natural to begin by clarifying the extent to which the two models place primacy on the actors.

The TERGM, in its basic formulation, is a model of the edges in a network. These edges are modeled as a function of the topology of the network itself, exogenous relational covariates, and exogenous vertex (actor) attributes, either concurrently or as a function of past states of the network (see Equation~\ref{tergm_t}). As a consequence, the TERGM assumes little in the way of actor agency; the TERGM is consistent with more detailed models that provide for high degrees of agency from the actors, but it is also consistent with models that assume no agency at all. In other words, the TERGM has little to say \emph{by virtue of its basic mathematics} about the primacy of actors or their agency, but it can be adjusted, through the dependence statistics and covariates included in its specification, to model the structure of the network as a function of actor-centric processes.

For example, conflicts in the international system can be modeled by describing structural effects like two-stars (an edge- or subgraph-level interpretation), but these counts of subgraph products can be interpreted from an individual actor perspective. For instance, one could posit that one state attacks another state because that state has been attacked by a third party before (preferential attachment), thus leading to a topology where two-stars are prevalent. The TERGM is therefore compatible with actor-based theories, but it does not assume per se that actors make these decisions. For instance, one could also apply a TERGM to the changing topology of electrical power networks and model endogenous features like two-paths without invoking theory related to the primacy of actors.

The SAOM is explicitly actor-centric, thus its namesake as an ``actor-oriented'' model, in the sense that one might regard the determination of outgoing edges as part of the behavior of the actor. The SAOM includes two processes, related to the timing and nature of edge changes, that are built specifically around the agency assumed of the actors \citep{snijders2010introduction} (see Equations~\ref{rate_function} and~\ref{changeprob}). Logically, these stochastic processes assume the agency of the actor: that the actor has the ability and motivation to change their edges. We will discuss these two processes again below, as they relate more to timing and edge updating than actors. What is important to make clear here is that the SAOM is not actor-oriented in the sense that the outcome of interest is an actor-level variable. Rather, similar to the TERGM, edges and the network topology resulting from these edges are the outcomes of interest. The SAOM is a model of the \emph{changes} in relationships between actors (thus, a model of the network and not a model of the actors) that is conditioned by specific assumptions about how actors behave.\footnote{We consider the original formulation of the SAOM as a model of network change here. For a discussion on SAOM variants pertaining to the behavior of actors, see Section~\ref{behavior}.} The TERGM is a model of the \emph{status} (e.\,g., connected or not) of relationships between actors \citep{block2018change} that may be adjusted to address changes in relationships.

It is also important to note that both models can contain identical sets of model terms in their respective $\textbf{h}$ vectors. It may \emph{feel} more intuitive to use the SAOM if one tests a theory that is based on methodological individualism. Yet, both models permit identical theoretical processes to be modeled, and any differences in theory are either related merely to a subjective interpretation one attaches to the equations, or they are related to the mini-step updating process of the SAOM that is absent in the TERGM \citep{block2018forms}. Therefore the remaining comparison of the two models will focus on the mini-step updating process inherent in the SAOM and absent in the TERGM.

\subsubsection*{Temporal Updating Process, Sequentiality, and Multi-Party Events}
For both the TERGM and SAOM, the network of interest is observed at several distinct, discrete time points, but one does not observe what happened in the network between observations. For example, assume a friendship network is sampled from a school on the last day of the school-year for four years. We observe four discrete networks, but do not know how the network changed between points in time. Some friendship edges observed at time $t$ may have dissolved by time $t+1$, and others may have dissolved and re-formed during this time. Depending on how often network edges change and how often the network is observed---properties that will vary greatly from application to application---, quite a bit of change may go unobserved between network snapshots. Often, one models networks at distinct time points because this is more convenient or cost-efficient than observing data on a continuous basis. In these cases, the network at time $t$ usually depends increasingly on the network at $t-1$ the shorter the time span between the observed time points.

The SAOM deals with this variability in temporal dependence by determining the required number of mini-steps as a function of the amount of change between the two consecutively observed networks \citep[147]{amati2011new}. That is, the simulations stop when as many edges have been changed as between the observed networks (with unconditional estimation) or after a pre-defined number of mini-steps reflecting the amount of change between the first and the second network (with conditional estimation) \citep{Snijders:2001}.

The TERGM, in contrast, is not built around the idea of modeling plausible paths between the observed time steps. It rather models only the data that are observed. In its simplest version, the TERGM is merely a pooled ERGM without any temporal dependence, thus assuming independence between consecutive networks. In most temporal settings, independence between time steps is unrealistic, therefore the user needs to think about how the previous network influences the current network. Instead of modeling this transition using mini-steps as in the SAOM, however, the outcome of this transition is mapped into the vector of user-supplied network statistics when the TERGM is used.

For example, suppose that preferential attachment shall be tested. In a SAOM, one would test for two-stars or indegree popularity over the (unobserved) sequence of mini-steps to establish whether preferential attachment is a plausible mechanism given the two observed time points. In a TERGM, one would test either for two-stars within the second time step or for two-stars over time (i.\,e., where the first edge is observed at the first and the second edge is observed at the second time step). Only in the second variant would the TERGM assume that the second edge occurred after the first edge of the respective two-star. In the first variant, simultaneous formation of both edges is equally plausible as a temporal order of the two edges.

One may be tempted to argue, based on this comparison, that the SAOM is better able to operationalize the concept of preferential attachment here because it posits a more fine-grained temporal updating process. Yet, this requires that the mini-steps accurately reflect the true temporal development on average, which will be tested below. Another way of putting it is that the SAOM creates artificial data to test a micro-level theory, but at the end of the day these artificial data can only encode information that is actually observed in the two observed time steps, therefore it cannot completely rule out that the edges are actually formed simultaneously or in reverse order (that is, in a way that is accounted for in the seemingly less specific TERGM).

This prompts an interesting philosophical question on the temporal order of social processes. One may argue that true simultaneity never occurs in physical systems and consequently does not occur in social systems either. Yet, a distinction needs to be made between absolute time and measurable time from the point of view of the entities that cause edge formation. For instance, in an e-mail communication network, sending an e-mail to two recipients on a single press of a button occurs at two consecutive time points because the server does not process the two recipients truly simultaneously. But from a theory perspective, this fine-grained temporal order does not matter because the sender does not make two separate decisions, but rather one joint decision to send out two e-mails, and the two e-mails are visible to the recipients (and possibly others) with such a small time interval between them that they are given the chance to respond to these e-mails as if they had arrived simultaneously.

An area of application in which simultaneity as perceived by actors matters is collective action theory. While true simultaneity is rarely achieved, theory assumes that groups collectively engage in non-action until the incentive structures are changed such that all actors have an incentive to become active at once. In network terms, it matters from a theoretical perspective whether the dissolution of an edge instantaneously creates an incentive for a group of actors to form edges or whether the dissolution of an edge leads to a single new edge somewhere in the network, that edge creates an incentive for somebody else to create another edge, and so on in a Markovian updating process. The SAOM assumes such a sequentiality in the temporal updating process while the TERGM remains agnostic as to whether edges were formed in a sequence or simultaneously as long as the statistics are found in the outcome network.

As a consequence, one can expect theories that posit simultaneity of multi-party events---such as theories building on collective action---to be more compatible with the TERGM while theories positing sequential edge updating should be equally compatible with both models. This should lead to different parameter estimates in the former but not in the latter case. The difference is not due to the sequentiality of the SAOM edge updating process per se, but rather due to the fact that the rate-of-change function may impose quite a long waiting time between edge events that are supposed to be simultaneous, thereby changing the network to some extent between supposedly simultaneous moves.

As a case in point, consider the network of international defensive alliances between, say, 1946 and 2000 with annually observed time slices, where NATO was formed in a single, coordinated treaty in 1949. Had opponents of NATO been allowed to react between the individual formation steps of the prospective NATO members (as would be possible in the SAOM), this may have deterred the founding states from completing the formation of NATO. The SAOM inevitably takes account of interactions between states because of the mini-steps. In the TERGM, such sequentiality can be avoided by incorporating non-temporal statistics for triadic closure within any given annual network, such as within 1949. In short, we expect the mini-step sequences of the SAOM to depart from the true data-generating process in applications that are based on collective action situations and other simultaneous edge formation or dissolution behavior. In other applications---perhaps friendship networks---sequentiality may or may not be a plausible assumption.

It is important to acknowledge that the expected difference is merely due to the temporal updating process assumed by the SAOM (and not assumed by the TERGM), not due to any user-supplied statistics, which can be identical. On the other hand, however, because the assumptions made by the TERGM are less specific, the analyst may face an increased burden in the specification of the model because she may need to add sufficient structure to the TERGM for it to match her hypothesized data-generating process.

\subsubsection*{Conditionality on Previous Time Points and Goodness of Fit}
There are two principal ways in which the SAOM and TERGM differ with respect to changes in the configuration of the network between observed time points. The first is simplest. The SAOM takes the first network to be given (not modeled) and then models changes from that base network going forward \citep{Lusher:2013}. It is natively a model of \emph{changes} in networks. The TERGM, which can be said to follow the core ERGM more closely, is a model of the network itself, rather than changes to it. The TERGM conditions on at least one previous network but natively models the structure of the network over time, rather than changes to it. That said, the TERGM can be adjusted to closely mimic the SAOM in the modeling of changes by including a ``memory'' term, which indicates whether a given edge existed in the previous time period (for an extended discussion of memory terms, see
\citealt{Leifeld:2018}).
In the TERGM, one can control for the effect of the previous network on the current network and then proceed by interpreting the parameters of the remaining statistics as network effects conditional on the previous state of the network.\footnote{In addition to an autoregressive memory term, the TERGM can include statistics that model the evolution between two or more consecutive time points as a process, for example delayed reciprocity over two time points or triadic closure over three time points. Thus, TERGMs need not be merely autoregressive but can model temporal change directly. An autoregressive TERGM without additional temporal dependence is a simple specification that may or may not be a good choice in a given empirical context.} The SAOM conditions on the previous network by letting the mini-step process depart from the previous network. While the amount of network change in the TERGM is visible in the parameter of the model term that conditions on the previous network, the amount of change in the SAOM is visible in the rate parameter(s).

Some may see the inclusion of such a memory term as advantageous because it makes the autoregressive process more transparent, whereas this process is obscured somewhat by the SAOM's updating procedure. Others may find the continuous-time structure of the SAOM advantageous for modeling a process that unfolds in continuous time (or nearly so) and is observed ``as snapshots'' at some moments, especially considering that the population parameters of the TERGM are sensitive to the interval of observation \citep{block2018change}. Ultimately, these are merely differences in the interpretation of the parameters. If the TERGM employs a memory term for conditioning on the previous network in lieu of the rate parameter of the SAOM, these differences in interpretation should not affect the fit of the remaining parameters in a given model if the same $\textrm{h}$ statistics are used to construct the model.

The two models also differ with respect to the role of observation frequencies or inter-observation times. Because the SAOM provides a data-generating process as part of the model, whereas the TERGM simply conditions on a previous network or networks, the parameter values of the SAOM that are not related to the rate-of-change function do not depend on the observation times or observation frequencies. TERGM parameters for model terms that posit dependence between time steps depend on the inter-observation times, potentially strongly. For example, consider a time series of yearly observations for 20 years. Modeling the series at yearly intervals with the TERGM will give different results for the temporal dependency terms than modeling the series at intervals of two or five years because the population parameters are different, not just because there are fewer data and because of random variability (which would also affect results). For the SAOM, the population parameters are invariant, given the rate parameter. As such, changing the period of observation for a SAOM would only affect the parameters through sparser data and random noise.

While independence of the time range between two observation points may be seen as an advantage in the interpretation of effects in the SAOM \citep{block2018change}, the flipside of this argument is that the role of the previous network(s) can be interpreted in a more transparent way in the TERGM precisely because the temporal dependency statistics (e.\,g., delayed reciprocity or memory terms) reflect the degree of dependence on previous time points in the parameter estimates.

For example, in a sequence of consecutive snapshots of the network of bilateral treaties between countries, there is usually little annual change. The TERGM makes this fact transparent by capturing this temporal dependence in a memory term (and possibly additional temporal statistics like temporal triadic closure). The remaining statistics do not conflate the previous state of the network with the respective statistic (e.\,g., triadic closure).

The difference between the two models has implications for goodness of fit: If the inter-observation times are very small, there is little change between networks, and the temporal statistics in the TERGM may indicate an artificially good model fit. However, it may be in the interest of the analyst to quantify the dependence on previous network states as part of the data-generating process, and hence the dependence on previous network states should be reflected in the quality of the model, depending on one's goal. In the TERGM, these model terms give an accurate representation of the degree of dependence on the previous time point(s) while the importance of the previous time point in the SAOM can only be evaluated by interpreting the rate parameters. As the role of the previous network for the current network is not entirely clear from the SAOM estimates, there is a risk of misattributing variation in the network to $h$ statistics that would be more easily explained by path dependency. Therefore one may argue that the TERGM does not \emph{artificially} inflate model fit, but rightfully attributes importance to the role of the previous network(s) rather than the updating process.

\subsubsection*{Longer-Term Temporal Dependence}
The TERGM assumes that temporal dependence occurs up to $K$ time steps in the past, and it requires the user to specify these dependencies. For example, it may be realistic that wars five to ten years ago impact wars in the current year, in addition to the wars just in the previous year. Depending on the application, the need to specify the dynamic process for the TERGM may raise the burden of specification on the analyst.

The SAOM assumes that dependence on events before $t - 1$ is completely factored into the previously observed network ($t - 1$). That is, it assumes a first-order Markov process. This places a lighter burden on the modeler with regard to model specification, but there may be cases where events that happened multiple time steps ago may not be captured by controlling for events at the previous time step.

As
\cite{Desmarais:2012}
note, these different specifications result in subtly but importantly different distributions assumed by the two models: TERGMs take the first $k$ networks to be given and do not model them, thus the networks that are modeled (those occurring after $k$ periods) are assumed to be in their stationary distribution conditional on the first $k$. The SAOM takes the first network as given, and models changes thereafter, thus following a stationary stochastic change process rather than a stationary stochastic series of graphs.

To make the contrast in basic assumptions clearer, consider some specific conditions for the two models. For the SAOM, the usual procedure for generating a sample is the updating process described in Section~\ref{saom}, whereas for the TERGM samples are usually generated through the indefinitely continued edge-wise updating of a Gibbs sampler. For the SAOM, sufficient conditions are its stochastic updating process (the usual procedure for generating a sample), and (more restrictive) myopic stochastic optimization by actors. For the TERGM, sufficient conditions are indefinitely continued edge-wise updating according to Gibbs sampling by edge variables (the usual procedure for generating a sample), and indefinitely continued myopic stochastic optimization by edge variables.

\subsubsection*{Actor Homogeneity, Strategic Behavior, and Size Limitations}
Both models assume that there is a homogeneous process operating on the network. More specifically, the SAOM posits that all actors have the same objective function and thus exhibit the same social behavior (on average): they wish to maximize their values of the network statistics included on the right-hand side of the specification. The TERGM makes a similar assumption by positing a single joint data-generating process over all dyads in the network. The differences only affect the interpretation, but not the underlying mathematics: while both models assume homogeneity of the process over the whole network, this homogeneity does not necessarily have to be interpreted as actor homogeneity in the TERGM.

In the TERGM, the user models counts of subgraph products. As pointed out in the subsection on temporal updating above, the TERGM is compatible with an actor perspective, but this perspective needs to be operationalized into countable edge-based network statistics \citep[see also][for a similar argument on REMs]{butts2017comment}. Therefore, rather than assuming that all actors have the same goals as in the SAOM, the TERGM assumes that the distribution of subgraph counts captures potential heterogeneity between vertices, thereby deviating from an actor-oriented perspective when necessary.

However, the SAOM can accomplish exactly the same task because one can specify identical subgraph products (read: $\textrm{h}$ statistics) in both models. For example, if one models the number of edge-wise shared partners per dyad in a TERGM, the distribution of edge-wise shared partners per dyad reflects heterogeneity among actors or dyads, but on average each dyad has the same probability to be located at a certain point in this distribution, which is captured by the estimate for the respective statistic. Similarly, the SAOM estimate posits the same tendency of actors to engage in edge-related behavior leading to the same outcome distribution of the statistic, and this distribution captures heterogeneity in the network in the same way as in the TERGM.

That said, in both models it is possible to model actor heterogeneity explicitly by adding vertex-level covariates (e.\,g., actor type) and interacting them with other elements of the $\bm{\theta}(N)$ vector. In this case, however, one defines exogenously where the heterogeneity is located in the graph.

The actor homogeneity assumption built into the SAOM has two consequences. First, the SAOM and the TERGM differ subtly with regard to their applicability to rational-choice theories. Many theories in the social sciences assume rational or otherwise forward-looking actors (e.\,g., game-theoretical models). In the SAOM, when actors evaluate their objective function at a given mini-step, they cannot ``look down the game tree.'' That is, the SAOM assumes that actors compare the network with singly permuted (one-edge changed) networks, not the long-run change that will result from immediate changes. As such, the updating process posited by the SAOM is theoretically incompatible with most rational-choice theories. Future research may try to tackle this problem by proposing SAOM variants that use a multinomial choice model with alternatives being evaluated by backward induction of user-specified games in lieu of Equation~\ref{changeprob}. Right now, the mini-step simulation process is likely led astray in many empirical applications because actors do not evaluate the responses of other actors to their actions. This may lead to omitted variable bias by design in situations where strategic actions matter.

The TERGM does not have provisions for strategic behavior built into the model either and thus potentially suffers from the same problem. Yet it is unclear whether the problem is equally severe in both models since the TERGM does not have a mini-step updating process, but rather models the probability of observing a specific network configuration. A potential solution may be to compute the equilibrium of the game one wants to model and including it as a network statistic. This of course suggests that one does not model the strategic interaction at the micro-level (i.\,e., every interaction in the game), but rather tests whether the outcome of this local process is present on average.

Neither of the two models has provisions for these kinds of extensions at this point, but they may be easier to implement in the TERGM from a user perspective because only the $\bm{\theta}(N)$ vector needs to be adjusted, not the temporal updating process built into the statistical model. In principle, the same equilibrium statistics could be included in the SAOM, but the interpretation would differ: theoretically, this would mean positing that actors homogeneously work together towards achieving the equilibrium, even though the equilibrium may actually be the result of a non-cooperative process that crucially rests on the assumption of heterogeneity of preferences. Therefore, while both models may or may not be able to arrive at the same estimates in game-theoretic situations, the usual actor-level interpretation of the SAOM may be confusing when applied to strategic situations while the TERGM does not have to be interpreted from a micro-level actor perspective. It is also unknown to what extent forward-looking behavior is prevalent in social networks, and it may vary considerably by subject area.

Finally, the homogeneity assumption may have consequences for the size of networks that can be modeled with each technique. The updating process of the SAOM assumes that all actors, when they have a chance to update their edge profile, observe the entire topology of the network in order to make their updating decisions. While this is reasonable in very small groups, it is hard to imagine that actors can know the entire topology of even medium-sized networks. Given that there are $2{{n \choose 2}}$ possible directed connections in a network, where $n$ is the number of vertices, one is tempted to think that the assumption of actors knowing the network topography is only valid for very small networks. Indeed, if there are only 150 actors in a network (e.\,g., a very small school), that implies 11,175 edges (or lack thereof) that an actor would have to be aware of and properly account for in order to use the objective function to update his/her edge profile.

While neither model is computationally capable of estimating models on very large networks, both can estimate models with thousands of vertices. However, the reasonableness of the SAOM assumption seems to hold only for small networks, thereby imposing an unspecified size limitation by theory. Users can estimate a SAOM even if this theoretical assumption is not met, but it is yet unclear if this leads to biased estimates because the Markov simulations are based on an unrealistic process. In the TERGM, in contrast, there is no theory that specifies such a size limitation. For example, if the user specifies a preferential attachment mechanism by including temporally delayed two-stars, this statistic is counted across the whole network without assuming that actors consider all other actors' edges and non-edges. I.\,e., the TERGM cannot depart from a theoretically specified Markovian updating process because it is agnostic as to whether such a process caused the transition from $t - 1$ to $t$. In the SAOM, the specification of such a model term assumes that actors make a choice over all other actors and their inter-connections in the network and thereby rely on perfect information.

What happens if the SAOM is applied to an empirical case where individuals cannot observe all other actors' relations? Future research will need to evaluate whether a violation of these theoretical assumptions in the respective models leads to biased estimates or whether these are merely differences in interpretation. While we cannot offer a comprehensive and nuanced comparison that takes into account all the differences pointed out in this review, we seek to illustrate the need for more focused comparisons below by contrasting two extreme cases that are maximally compatible with the respective assumptions of the SAOM and the TERGM and by demonstrating that the plausibility of a model's theoretical assumptions is consequential for its performance.

\subsubsection*{Behavioral Component and Estimation} \label{behavior}
One of the great innovations of the SAOM, and an often-cited reason for using it \citep{veenstra2013network}, is that it includes the option to model a dynamic behavior (vertex attribute) ``simultaneously'' with the evolution of the network.
\cite{Desmarais:2012}
showed that the TERGM can also incorporate a behavioral component, though the dynamics of the two work somewhat differently. Both behavioral components are statistical models for the vertex attribute at time $t$ that include, on their right hand sides, other vertex attributes and vertex-level measurements on the network (e.\,g., the centrality of the vertex).

Where the two behavioral components differ is that the SAOM iteratively estimates the behavioral model for each of the simulations between time periods conditioning the behavioral model on the most recent simulation values available. The TERGM simply employs a (generalized) linear regression and conditions on the previous time point. It is also the case that, if actor behavior is being modeled as part of the network-generating process as in the SAOM, simulated intermediate vertex behavior will also be accounted for in the network-generating model \citep{Steglich:2010}. Though the iterative simulations can imply a more immediate effect of previously observed networks on behavior, these models do not incorporate any simultaneous dependence between the network and behavior.

Thus, it is fair to characterize the behavioral component of the SAOM as more dynamic than that included in the TERGM. However, it is also unclear \emph{a priori} which formulation will perform better in explaining and predicting vertex attributes. Whether or not a behavioral model that incorporates simulated intermediary steps rather than simply conditioning on observed behavior is more realistic will be entirely dependent on the accuracy of the simulation process. If the edge-updating process described above is out of step with the data-generating process, the more dynamic process may actually perform worse than the simpler behavioral model.

We leave the comparison and evaluation of model components for vertex attributes to future research and focus on the network aspects of the two models. Both models have these capabilities, but like the network parts, they differ substantially in their assumed updating processes.

Moreover, as outlined above, the two models have different estimation strategies and each may be estimated in several different ways---the SAOM is typically estimated via generalized method of moments but can also be estimated with Bayesian techniques, and the TERGM is usually estimated by MCMC-MLE for smaller networks and by bootstrap-corrected maximum pseudolikelihood for larger networks. Though each of the estimation procedures is different and may require care to different things (e.\,g., good mixing and convergence with Markov chains), they should all provide valid estimates of any given model. That is to say, any valid estimation strategy must produce unbiased and consistent estimates. Any estimation strategy that is not unbiased and consistent is not statistically sound and should not be used as it will not produce valid results. We see the topic of estimation as somewhat beside the point of the present analysis: unless any of the approaches are outright wrong---and each has a substantial literature behind it---then it should not matter which particular strategy is used. Put another way, if we are willing to assume the validity of each of these estimation strategies, any differences in fit and performance between a TERGM and a SAOM with comparable specifications will be the result of the model structure.

The TERGM and SAOM do, however, share a common limitation with respect to estimation: estimation of both models is computationally intensive and time-consuming. As a result, both models tend not to be estimable in short time periods (weeks/months) for very large networks. As increases in computational power meet efficiencies in coding and advances in parallelization, this restriction is weakening over time. However, as of this writing, neither model is easy to estimate as the number of vertices and temporal observations increase.

\section{Simulation-Based Comparison}

\subsection{Theoretical Expectations}
The previous section considered a number of differences in the theoretical assumptions of the TERGM and the SAOM, despite the many similarities of the two models. It is yet unclear in how far these theoretical differences also translate into a difference between estimated parameters and the parameters of the true data-generating process when the theoretical assumptions are not met by the data.

Here, we select two critical cases for comparison, one with a data-generating process (DGP) that is maximally compatible with the SAOM and one that is maximally compatible with the TERGM.\footnote{Note that one might also approach this question by conducting simulation experiments in order to test each of the identified differences for an impact on model performance. For example, one could test whether increasing levels of edge simultaneity in the data-generating process lead to a deterioration of model performance in the SAOM and in the TERGM, respectively, all else being equal. Our theoretical comparison suggests that both models should fit equally well if all edges are sequentially established while the TERGM should have a relative advantage with increasing prevalence of multi-party edge events or incentives for collective action. However, so many separate simulation experiments would far exceed the page length boundaries afforded by a journal article.} Our expectation from the theoretical comparison is that the performance of the respective model is better when the underlying DGP of the simulation is compatible with the assumptions of the respective model.

If a SAOM DGP (i.\,e., a rate and an objective function) is used to simulate a sequence of networks, the SAOM should show the same or potentially a better fit than the TERGM. If a TERGM DGP is used to simulate a sequence of networks, which may violate some of the more specific assumptions of the updating process in the SAOM, the TERGM should exhibit a better model fit than the SAOM. As the properties of the DGP are modified to match the assumptions of the SAOM, the difference in goodness of fit should disappear.

More specifically, for the SAOM DGP, we expect that a Markovian data-generating process without strategic action in a moderately-sized network without simultaneous edge formation should be recovered more adequately by the SAOM because such a process meets all theoretical assumptions of the SAOM and its updating process is geared specifically to such a DGP. The TERGM is also compatible with these assumptions but may be less precise as the it would also be compatible with other DGPs. We simulate such a process using the \texttt{NetSim} package \citep{stadtfeld2015netsim} for the statistical programming environment \texttt{R} \citep{coreteam2016r:}.

We also expect that a TERGM DGP determined by discrete updating by actors and from an edge-based perspective using a Metropolis-Hastings network sampler as implemented in the \texttt{ergm} package \citep{hunter2008ergm}, rather than an imposed continuous-time sequential edge updating process by actors, should lead to a better model fit when the TERGM is estimated compared to the SAOM because some of the specific assumptions of the SAOM may not be met.

This comparison of two extreme cases---one directly using the data generating process of the SAOM and the other of the TERGM---serves to illustrate that these differences do matter and that one should pay careful attention to select an appropriate statistical technique in applied research. If we do find differences between the two models based on the two different processes, this may suggest that empirical networks may also vary on this continuum or even beyond, and this will pave the way for a research agenda where more focused comparisons can be conducted.

Further, note that while our goal in the theoretical discussion above was to discuss the TERGM and SAOM independently of their software implementations, software necessarily plays more of a role in this performance-based exercise. A model is certainly not equal to its software implementation, but the constraints and options of the software do affect the performance of the model on real data. We try, in what follows, to be as transparent as possible about what is a software limitation and what is a model limitation.

\subsection{Out-of-Sample Prediction}
We select the most impartial criterion for inter-model comparison possible: out-of-sample predictive performance. Out-of-sample prediction involves the fitting of a model on one dataset, called the training set, and the application of that fitted model to a different dataset, called the test set. This is a powerful framework for performance testing because, as long as there is no overlap between the training and test sets, the only thing they will have in common is the data-generating process (DGP). Out-of-sample prediction is an impartial criterion because (a) it is easily comparable across methodologies, whereas other fit criteria are not; (b) in-sample prediction runs the risk of overfitting, where the parameters of the models might capture idiosyncrasies of the sample rather than the DGP;\footnote{A model that is overfitted will predict well in-sample (e.\,g., on the training set), but poorly out-of-sample (e.\,g., on the test set).} and (c) out-of-sample prediction ensures that the model that best captures the DGP in nature will perform best. Though it is generally possible that well-performing predictive models do not correspond to meaningful explanatory theories, here we use it as a criteria to distinguish between two models that have identical right-hand-side specifications when the true data generating process is known. We see this as uncontoversial because it would be odd to claim that, between two models with identical predictors, the model producing the worst fit should be qualitatively superior.

Much of what follows may seem evident from other areas of statistics. However, as there is little published work detailing the use of out-of-sample prediction with these longitudinal network models, we take the time to consider some of the details of this methodology. In longitudinal network models, a training set of older networks is used to predict a test set of newer networks in the time series. Prior applications of out-of-sample prediction to SAOMs include
\citet{Desmarais:2012},
\citet{kinne2013network}, \citet{koskinen2012modelling}, and \citet{warren2016modeling}. There is also a substantial literature looking at out-of-sample prediction with other network models \citep[e.\,g.,][]{chiba2015every, montgomery2015calibrating, ward2013learning}.

\subsubsection*{Temporal Heterogeneity}
One may object that the DGP may change between the training and test sets, rendering the predictive exercise meaningless. If the DGP changes substantially between the training set and the test sets, one should indeed expect to see poor out-of-sample predictive performance. However, it is also the case that the very goal of longitudinal network models is to describe a joint process taking place across multiple time steps. As such, if the longitudinal network model is applied correctly, such out-of-sample testing should not be problematic because the DGP should not be changing. If the DGP were changing, the use of a longitudinal network model, either the SAOM or the TERGM, would be problematic. If it is sensible to assume that the same model can be applied to all waves of a network, then it should also be sensible to estimate the same model on the first $k$ waves and test it on the $(k+1)$\textsuperscript{th} wave using out-of-sample prediction since the data-generating process does not change. Relatedly, if the model performs well out-of-sample, this is strong evidence that the DGP has not changed.

\subsubsection*{Prediction or Explanation?}
Ultimately, most analyses are interested in explaining network phenomena rather than predicting them \citep{block2018forms}. It might seem then that a predictive exercise such as ours is assessing the model on the wrong criteria. We disagree, however. Though we use our models to predict, they are not specified atheoretically such that their specification is unrelated to an explanatory model. Any model specified to explain data from a specific DGP should be applicable to new data that emerge from the same DGP. As long as the DGP is not expected to change, as discussed above, testing an explanatory model out of sample is a powerful means by which to judge the extent to which it actually captures the DGP.

\subsubsection*{Predict Network Structure or Edge Probability?}
It is not immediately clear whether a predictive exercise for network models should be attempting to predict the frequency of occurrence of certain network structures (e.\,g., closed triads) \citep{hunter2008goodness} or predict the formation/persistence of dyadic edges. Our solution to this is to examine predictive performance on both, endogenous network characteristics and the location of edges in the graph. As the literature does not host much discussion on this topic, this needs elaboration: Though it may seem very natural to attempt the prediction of edges---indeed, such would be the typical approach in statistics and the voluminous literature in computer science---, there is a case to be made against this approach. Consider an extreme case where the network is generated only by endogenous network structure and not at all by exogenous covariates. In such a case, the labels of the vertices should not matter and therefore edge prediction should not be informative. That is, one would not expect the predictive performance of the model, at the edge level, to differ from random, but one may be able to predict structures, such as the number of closed triads, with substantial accuracy. This is the reason that software packages such as \texttt{statnet} \citep{handcock2008statnet:, hunter2008ergm} and \texttt{RSiena} \citep{ripley2017manual, snijders2010introduction} focus their goodness-of-fit tests on network structures rather than dyadic edge predictions.

However, this extreme case is misleading as the order of vertex labels will matter in most empirical applications. On the one hand, most models will feature \emph{some} exogenous dyadic or vertex-level covariates, which makes it important to predict edges between those vertices where the variation occurs. On the other hand, the absolute location of edges and subgraph structures in the network is especially important in longitudinal network models, such as the ones considered here, because otherwise local structures cannot be carried forward from one time step to the next. It would be highly unrealistic to assume that structures are sticky and change only as a Markov process during the mini-step updating, but then suddenly change their location in the network completely when the next time period starts. Therefore most useful applications feature some degree of path-dependence of the absolute location of edges, and it is an important consideration \emph{who} is connected, rather than merely to what extent edges or certain structures show up \emph{somewhere} in the network. For this reason, both measures are important aspects of model fit: auxiliary statistics capturing certain aspects of network structure as a measure of the endogenous goodness of fit and the dyadic prediction performance as a measure of the extent to which edges are predicted in the right ``spot'' in the network.

Furthermore, as the theory becomes more and more about exogenous factors, the more useful edge prediction becomes. Consider the opposite extreme to that considered above: a scholar has a theory about how exogenous factors cause edges to form in a given network. He/she models this process with a TERGM or SAOM because he/she is worried about controlling for potential endogenous dependencies, even though they are not of primary theoretical interest. In such a case, the dyad-wise predictive accuracy of the model would be paramount, and the extent to which the model predicts structures would be important only to satisfy the scholar that he/she had appropriately modeled any endogenous processes that may be at work.

As such, we believe both types of goodness-of-fit assessment have merit and are actually complementary. It is further important to keep in mind that both types of assessment perform well with out-of-sample prediction (as well as in-sample prediction).

\subsection{Predictive Methodology} \label{predictivemethodology}
We generate the out-of-sample predictions in the following way. For each of the below applications, we predict the final observation of the network using only observations that occur temporally prior. So, for both the TERGM and the SAOM, the first network is not modeled  (it is used for temporal conditioning by both models) and the last network in the series is not modeled either because it is the object of prediction.

For example, in a network that is observed at four time points, a TERGM for $t=2$ and $t=3$ is estimated based on the corresponding covariates at $t=1$ and $t=2$, respectively. These covariates can be functions of the previous network, and in the cases we report below, we use a dyadic stability term that captures to what extent edges and non-edges are sticky over time. Once such models have been estimated, their resulting coefficients are used to simulate several new networks based on the covariates at $t=3$ using MCMC sampling, and the simulations created in lieu of time point $4$ are compared to the observed network at $t=4$. If the simulations predict the empirically observed network at time step $4$ well, we can conclude that the out-of-sample predictive performance of the TERGM is good.

The procedure for SAOMs is analogous to that of the TERGM: we let the SAOM simulate edge changes and estimate parameters between $t=1$ and $t=2$ and between $t=2$ and $t=3$. Then we extract the estimated parameters from the model and simulate a SAOM process forward from $t = 3$ to $t = 4$, holding the extracted parameters constant. This is repeated multiple times in order to yield multiple simulated networks. The last simulation mini-step between $t=3$ and $t=4$ (based on unconditional estimation) is used to predict the actual network at $t=4$. Both procedures, the SAOM and the TERGM out-of-sample prediction, are thus comparable.

The actual comparison between the observed network and the distribution of simulated networks for the same time step is carried out in two ways: first, we assess the goodness of fit by comparing several simulated network statistics to their observed counterparts, for example the distribution of geodesic distances or the distribution of edge-wise shared partners \citep{hunter2008goodness}. If these distributions are similar to the observed distributions, we can be confident that the network-generating process is captured by the model. We compare this fit across the different types of models (SAOM vs.\ TERGM). Second, we assess the classification performance of the simulations produced by each model using receiver operating characteristic (ROC) and precision-recall (PR) curves and by computing the area under the curve (AUC) for each of the two curves \citep{davis2006relationship, hanley1982meaning, sing2005rocr:}. In other words, we assess the extent to which dyadic states (edge or no edge between each $i$ and $j$ vertex) are predicted accurately by the simulations. More specifically, these procedures provide an intuitive understanding of sensitivity and specificity, or type I and type II error, or the true positive rate and the true negative rate, for the prediction of edges in the network.

\subsection{Evaluation of Predictive Fit using Simulated Data}

\subsubsection*{Setup of Simulation Experiments}
For our first test of the predictive performance of the SAOM and the TERGM, we create artificial network data under a known data-generating process and then fit the same model to the networks that was used to create the data in the first place. This does not constitute a universal test of the predictive performance of the two models because there is an infinite set of possible model specifications one could compare. Yet, if one of the models outperforms the other model on a relatively simple specification across a range of different parameters, this will be a strong indicator of differential predictive performance in general. We choose two simple data-generating processes that are maximally compatible with the TERGM and the SAOM, respectively. They are based on identical, simple endogenous network statistics and differ only in their updating processes.

\begin{figure}[t]
\begin{center}
\includegraphics[width=0.9\textwidth]{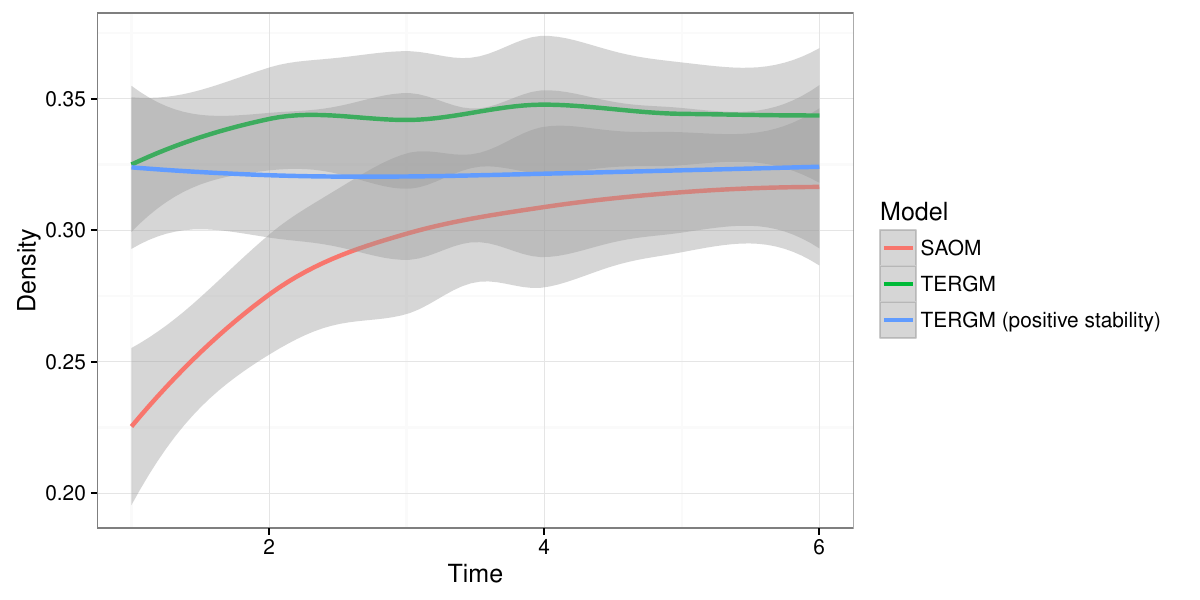}
\end{center}
\caption{Density of 100 simulated networks for the SAOM DGP and the TERGM DGP.}
\label{fig:simulation-densities}
\end{figure}

First, we independently draw four parameters from a uniform distribution ranging from $-3$ to $+3$ at steps of $0.001$. These are typical parameter values one could find in an empirical SAOM or TERGM. Naturally, some combinations of these parameters (e.\,g., extreme combinations where all four parameters are close to $-3$) may lead to degenerate simulations, i.\,e., full or empty networks. If we encounter such nearly full or empty networks, we drop the parameters and sample a new set of parameters until they do not yield degenerate networks anymore. In practice, we drop parameters if they lead to networks with a density smaller than 0.03 or larger than 0.97 at any point in the sequence. Figure~\ref{fig:simulation-densities} shows the mean densities of the simulated network sequences and their variation over the temporal sequence, after removal of degenerate specifications.

Second, the first three of the four parameters are used to simulate a series of six networks with $20$ vertices based on three model terms (and the three sampled parameters attached to them):
a baseline edges term
\begin{equation}
 \textrm{h}_{\textrm{edges}} = \sum_{i \neq j} N_{ij},
\end{equation}
(or density in the case of the SAOM), a reciprocity term,
\begin{equation}\label{reciprocity}
\textrm{h}_\textrm{r} = \sum_{i\neq j} N_{ij}N_{ji},
\end{equation}
and a transitive triplets statistic
\begin{equation}
 \textrm{h}_{\textrm{trans}} = \sum_{i,j,k} N_{ij}N_{ik}N_{jk}.
\end{equation}
Note that the $i$ index is removed from the sum in each case for the SAOM in order to account for the actor-oriented perspective.

One of the two simulation experiments---the TERGM process---uses MCMC and an ERGM formula (see Equation~\ref{ergm}) based on the three sampled parameters to create a single network. Based on this simulated network, a series of five new, consecutive networks (time steps $t = 2$ to $t = 6$) is simulated forward using the same three terms as for the first network and an additional dyadic stability term
\begin{equation}\label{stability}
 \textrm{h}_s = \sum_{ij} N_{ij}^{t} N_{ij}^{t-1} + (1- N_{ij}^{t}) (1-N_{ij}^{t-1})
\end{equation}
with the fourth sampled parameter value attached to it.

For the SAOM process, an empty network is created, and the SAOM actor selection and edge updating process is iterated five times to obtain networks from $t = 2$ to $t = 6$.
For the rate-of-change function (Equation~\ref{rate_function}), a parameter value between $5$ and $10$ is sampled uniformly, which means that each actor is expected to re-consider his or her local network configuration this number of times per time unit \citep{stadtfeld2015netsim}. This parameter choice is based on examples in the \texttt{RSiena} manual \citep{ripley2017manual}.

In both simulation experiments, the first network in the list is removed, and the five networks that are based on the same edges, reciprocity, and transitive triplets parameters but different treatments of the updating process are retained for further analysis.
In either case, the result of this procedure is a series of five networks which are serially correlated and based on several simple endogenous statistics.
100 such series of networks are independently created according to these rules in each experiment, and a SAOM and a TERGM are applied to each series of networks in each experiment.
Each of these models contains an edges, reciprocity, and a transitive triplets model term.
While the TERGM contains an additional edge stability term (see Equation~\ref{stability}) for modeling the temporal updating, the SAOM uses its rate-of-change function and the objective function to model change between the networks.
Thus these models should be able to capture the respective original data-generating process well---unless their inherent assumptions limit their ability to do so.
Moreover, the generality of each model can be assessed by cross-checking the model fit against the other data-generating process.

We simulate 100 sets of five consecutive networks for each procedure, estimate 100 SAOMs and 100 TERGMs, simulate 10 new networks from each of the 100 respective models out of sample, and then assess the predictive fit (as described in Section~\ref{predictivemethodology}).
In both cases, the simulations are compared against the actual last network from the series using the ROC and PR curves.
To quantify the predictive performance of each model, the area under the curve (AUC) is computed for both curves.
In each simulation experiment, this results in 100 AUC-ROC measures and 100 AUC-PR measures for the SAOM and the same quantities for the TERGM.

\subsubsection*{Results: Tie Prediction}
A two-sample $t$ test for ROC ($t(180) = 7.21,\ p < .001$) and PR ($t(196) = 4.85,\ p < .001$) reveals that the fit of the TERGM models is better than the fit of the SAOM on average in the TERGM experiment while the fit of the SAOM models is better than the fit of the TERGM on average in the SAOM experiment, though there is some statistical ambiguity for the SAOM DGP ($t(195) = -1.98,\ p = .049$ for ROC and $t(197) = -1.40,\ p = .162$ for PR).

\begin{figure}[t]
\begin{center}
\includegraphics[width=1.0\textwidth]{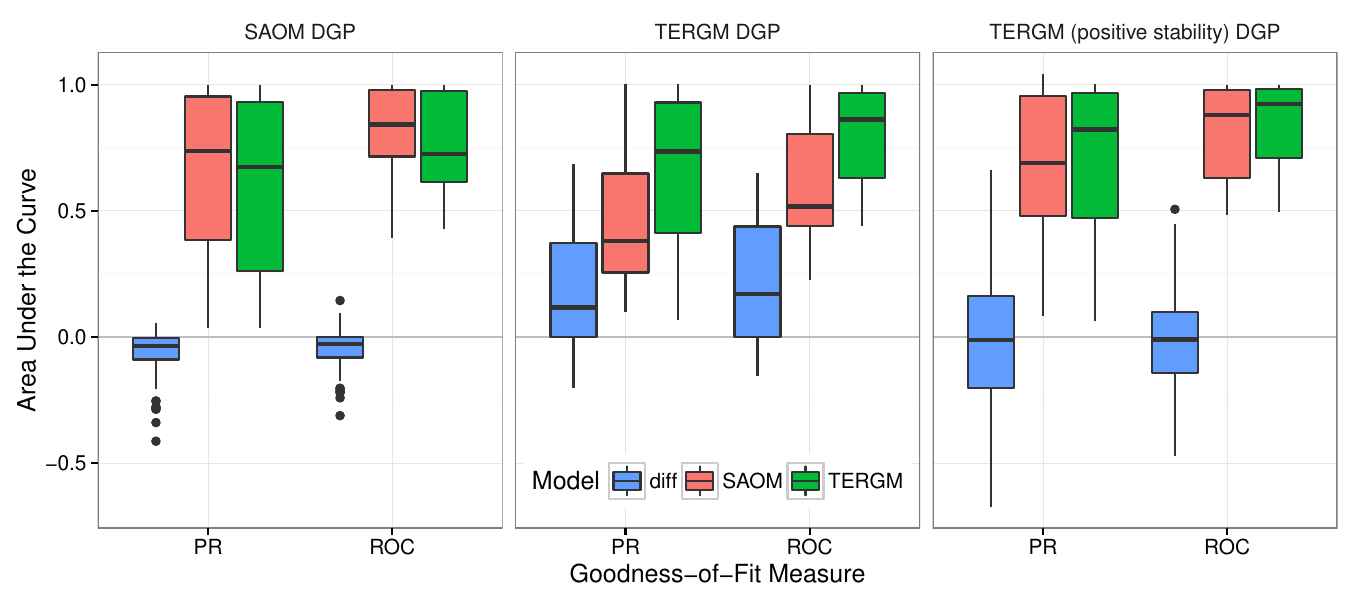}
\end{center}
\caption{Comparison of the area under the curve (AUC) for receiver operating characteristics (ROC) and precision--recall (PR) curves for SAOM and TERGM (with two specifications). Simulations from a data-generating process that is maximally compatible with the SAOM (on the left), the TERGM (in the middle), and a modified TERGM restricted to exhibit positive dyadic stability (on the right).}
\label{fig:simulation-auc}
\end{figure}

To see this more clearly, the left and middle panels of Figure~\ref{fig:simulation-auc} show boxplots that visualize the distribution of all four AUC measures per simulation experiment as well as the differences between models ($\textrm{diff}_s = \textrm{AUC}(\textrm{TERGM}_s) - \textrm{AUC}(\textrm{SAOM}_s)$, where $s$ denotes the simulation iteration). The results indicate that the TERGMs usually outperform the SAOMs in the TERGM DGP experiment, irrespective of whether non-edges are part of the performance measure (ROC) or not (PR), as demonstrated by the fact that most of the ``diff'' distribution is positive in both cases. There are, however, also a few cases where the SAOM fits better. Future research should investigate whether this is due to a low consistency of the estimator or due to a systematic pattern. In the SAOM DGP experiment, where the data-generating process of the SAOM is precisely modeled, the difference between the estimated models is slightly in favor of the SAOM, but the significance is ambiguous (see $t$-test above).

These two DGPs capture ideal-type conditions. If one deviates from the specific assumptions compatible with either model, the difference in model fit should disappear. We test for this possibility in a third specification by altering the TERGM DGP to include only positive parameter values for dyadic stability, as shown in the third panel in Figure~\ref{fig:simulation-auc}. Empirical networks for which the SAOM was designed, such as friendships, rest on the assumption of positive stability because friendships are ``sticky'' over time. Some other empirical networks, for which the SAOM was not originally designed, feature negative stability. Cases in point include 1.\ networks with a set of rules determining alternating edges (such as speed dating networks, membership nominations among groups of people, or sports tournaments, where individuals are not paired with people they have previously engaged with and vice-versa); 2.\ networks where a tie has corrective functions and becomes redundant when the tie has served its purpose (such as punitive conflict relationships or information transmission ties, where the tie becomes obsolete when the information has been transmitted); and 3.\ bipartite networks with a temporally changing or passive second mode (such as authorship networks of academics and joint papers, where an academic authors a paper only once at most, online purchase networks of buyers and books, where a reader buys a book only once at most, honey extraction networks of bees and flowers, where extraction leads to resource depletion, or sponsorship networks of legislators and bills, where a sponsorship tie is an instantaneous event and is not formed a second time in the next observed time period). The positive stability specification of the TERGM DGP (third panel of Figure~\ref{fig:simulation-auc}) deviates from such examples by restricting networks to be positively autoregressive and thus in better keeping with the SAOM assumptions. The advantage in tie prediction for the TERGM disappears when this aspect of the DGP is changed, with the AUC becoming indistinguishable between TERGM and SAOM ($t(196) = 1.30,\ p = .197$ for ROC and $t(196) = 0.459,\ p = .647$ for PR).

The key result here is that one or the other model may have an advantage depending on the specific nature of the DGP and the extent to which the assumptions of the network model match the nature of the DGP. This affects aspects such as positive or negative stability as in this illustration, sequential versus simultaneous edge formation, and other aspects covered in the theoretical discussion above.

\subsubsection*{Results: Endogenous Fit}
While the AUC comparison focuses on the actual location of edges in the network, another important feature is whether a model fits well in terms of its endogenous properties.
This aspect of the goodness of fit can be checked by comparing auxiliary statistics like the shared partner distribution, the degree distribution, and the distribution of geodesic distances between the originally observed network and the out-of-sample simulations of the model for the same time step.
In order to compare the 100 SAOMs and TERGMs per study, we plot the difference between the absolute deviation of the median TERGM simulation from the original network and the absolute deviation of the median SAOM from the original network.
If $a_s$ denotes the original observation of an endogenous statistic (for example, how many dyads in the network have exactly four dyadwise shared partners?) for the $s$'th simulation, $b_s$ denotes the median of same count for the $s$'th TERGM, and $c_s$ denotes the median for the SAOM, then
\begin{equation} \label{eq:diff}
 \textrm{diff}_{\textrm{endogenous}} = \frac{1}{100} \cdot \sum_s^{100} |c_s - a_s| - |b_s - a_s|
\end{equation}
is the difference in deviations from the original network for a given statistic.
Values above $0$ reflect a greater deviation from the true value for the SAOM, and values below $0$ reflect a greater deviation from the true value for the TERGM.

\begin{figure}[tp]
\begin{center}
\includegraphics[width=1.0\textwidth]{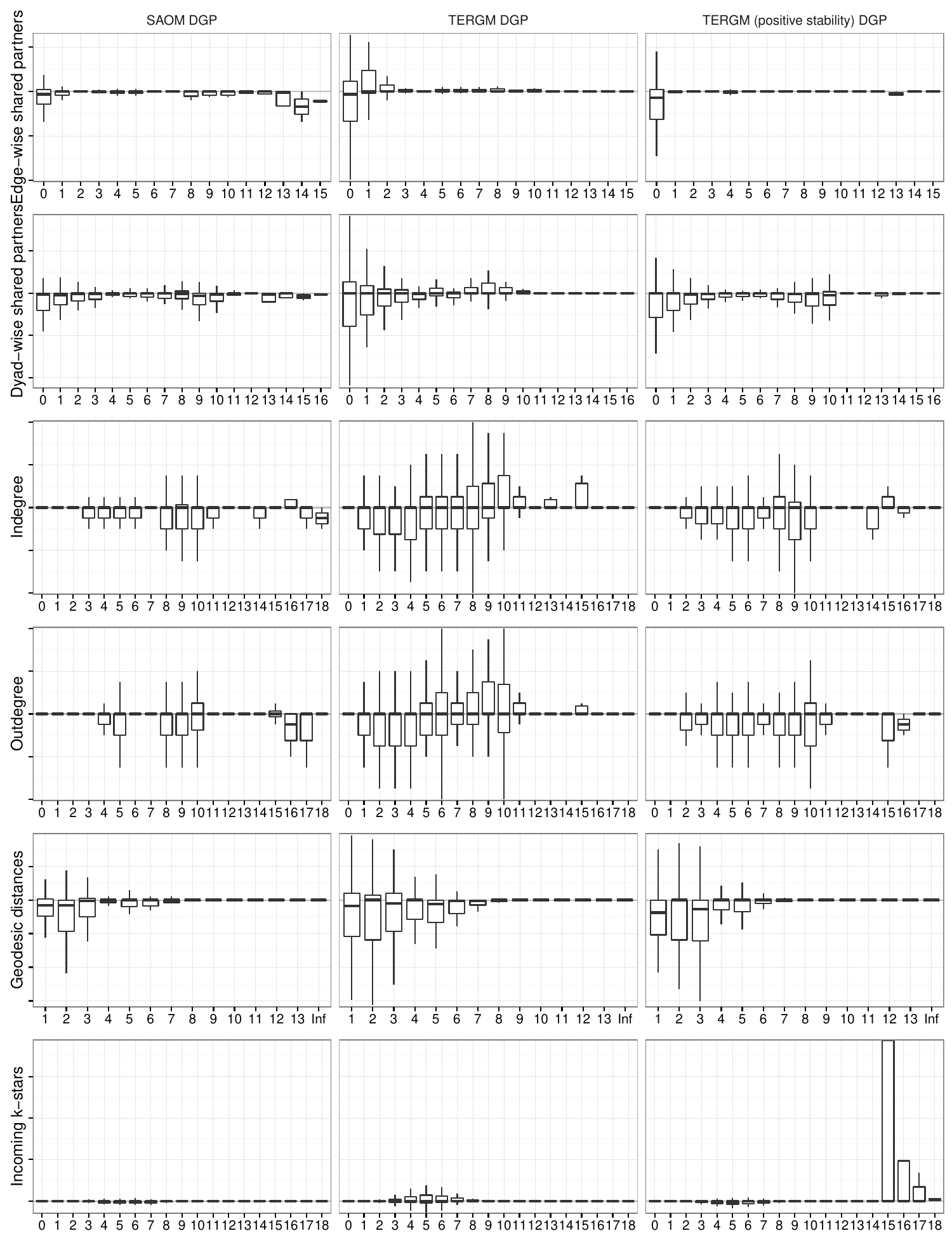}
\end{center}
\caption{Endogenous model fit. Simulations from a data-generating process that is maximally compatible with the respective model. Comparison of the SAOM and the TERGM for each DGP based on endogenous network properties. The boxplots show the difference between the absolute deviation of the TERGM from the original data and the absolute deviation of the SAOM from the original data. Distributions above 0 indicate a better performance of the TERGM, and distributions below 0 indicate a better performance of the SAOM.}
\label{fig:simulation-endogenous}
\end{figure}

Figure~\ref{fig:simulation-endogenous} uses boxplots to visualize the distributions of these values for the three DGPs.
There are no noticeable systematic differences between the two models across the three DGPs, with the medians usually being around zero, with a few exceptions.

The preliminary main finding here is that the two models seem to capture endogenous structure equally well irrespective of context. Thus it is fair to say that both models arrive at roughly the same conclusions regarding the topology of the network, while they differ in their assessment of which vertices are involved in the respective edges or structures, depending on the assumptions built into the model and their match with the respective DGP.

As a caveat, we should note that this comparison is only based on a subset of the theoretical differences enumerated above. For example, strategic action or the size limitation were not tested for detrimental effects on model performance, so we cannot rule out that endogenous performance differs as well in some applications. The next section introduces such a case of diverging endogenous fit in an empirical context.

Even if an exhaustive answer about predictive performance cannot be given due to the infinite set of possible models one could test, this simulation exercise is simple and plausible enough to provide a first hint that the two models may differ in a range of applications where the data-generating process is not perfectly in line with the specific assumptions of the respective model. This underlines the need for comparative goodness-of-fit assessment in empirical applications to justify model selection.

\section{Real-Data Example: Replication of a Friendship Network SAOM}
So far, we have established that there are theoretical differences between the updating processes of the TERGM and the SAOM, with a more specific process posited by the SAOM. Even in a simple simulation exercise, these differences have been shown to be consequential for the inferential performance of the respective model, at least with regard to edge prediction, but we do not know yet which specific aspects cause these differences. Out-of-sample prediction is a useful tool for comparing the performance of the two models on a given (simulated or empirical) dataset, though this certainly does not replace the role of theory in guiding model selection \citep[cf.][]{block2018forms}.

The remainder of this article is dedicated to these further points: Sometimes theory is not sufficient to evaluate a priori if the assumptions of one model are reasonable in a given context or if the other respective model should be used instead, as further research is needed on whether and how the theoretical differences translate into different empirical performance. For example, even in apparently well-studied empirical contexts (such as the example provided below), the specific updating posited by the SAOM may not be at work, leading to a better performance of the TERGM. Therefore, when in doubt, users can easily use our companion software to assess the out-of-sample performance of both models. This will contribute in the long run to our collective understanding of the contexts in which one model should be chosen over the other model. Moreover, we will demonstrate by example that empirical cases exist where endogenous model fit differs out of sample between the two models (in contrast to the simulations, where performance regarding edge prediction differed).

To make these points, we illustrate the performance-based comparison of the SAOM and the TERGM by replicating an analysis reported by \citet{snijders2010introduction}.
Substantively, the authors model a friendship network in a Dutch school class over several time steps.
The data were originally collected by \citet{knecht2006networks, knecht2008friendship} and later used as the primary example for introducing stochastic actor-oriented models \citep{snijders2010introduction}.
According to what we believe we know a priori about friendship networks, the SAOM should be well suited for modeling the evolution of friendship because the updating assumptions as set out above seem to be reasonable in this context.
The SAOM was originally designed with the application of friendship network evolution in mind, and friendship networks are still one of the primary fields of application.
Furthermore, the application we consider is a frequent example used in expository papers, workshops, and tutorials for the SAOM. The application has been studied extensively with the SAOM and is ``standard'' in its literature.
Therefore this empirical application serves as a critical test case that is maximally tailored to the assumptions of the SAOM.
For example, more specifically, one could argue that it is reasonable to think that actors do in fact change their relations one at a time without coordination and that actors will alter their edges to ``improve'' their position in the network.

The friendship networks under study were measured between September of 2003 and June of 2004.
The data consist of 26 students in their first year of secondary school and were recorded in four waves over the period of study.
The subjects, 17 girls and 9 boys aged 11--13, were asked to nominate up to twelve classmates they considered ``good friends.''

To address the fact that friendship networks tend to be reciprocal, sex-segregated, and show tendencies towards triadic closure, \cite{snijders2010introduction} include reciprocity (see Equation~\ref{reciprocity}), transitive triplets,
\begin{equation}
\textrm{h}_{\textrm{t-trip}} = \sum_{j,k} N_{ik}N_{ij}N_{jk},
\end{equation}
transitive edges,
\begin{equation}
\textrm{h}_{\textrm{t-edges}} = \sum_k N_{ik} \max_j (N_{ij}N_{jk}),
\end{equation}
three-cycles,
\begin{equation}
\textrm{h}_\textrm{c} = \sum_{j,k} N_{ij}N_{jk}N_{ki},
\end{equation}
and an indicator for when friendship edges are sexually homophilous, thus capturing sex segregation:
\begin{equation}
\textrm{h}_{\textrm{same-sex}} = \sum_{j} N_{ij}X_{ij}.
\end{equation}
Furthermore, they include the following degree-based measures as controls: indegree popularity
\begin{equation}
\textrm{h}_{\textrm{in-pop}} = \sum_j N_{ij} \sqrt{N_{+j}},
\end{equation}
out-degree popularity
\begin{equation}
\textrm{h}_{\textrm{out-pop}} = \sum_j N_{ij} \sqrt{N_{j+}},
\end{equation}
out-degree activity
\begin{equation}
\textrm{h}_{\textrm{out-act}} = N_{i+}^{1.5},
\end{equation}
as well as exogenous indicators for male respondents (as a sender effect and as a receiver effect) and friendship in primary school.

We replicate a specification used by \citet{snijders2010introduction} for this comparison. Table~\ref{tab:knecht} shows the results of a faithful re-analysis of their ``Model 0'', the most elaborate model without actor--behavior co-evolution reported by \citet{snijders2010introduction}, and a TERGM with the same specification. In other words, both models include effects for Equations~11 and 15--21. Our goal is not to make a theoretical contribution to the friendship literature or to find the best possible fit to these data. As a consequence, we retain the canonical and widely replicated specification used by \citet{snijders2010introduction} rather than formulating a novel specification.

While the model terms are faithful to the original analysis by \citet{snijders2010introduction}, the estimates reported in Table~\ref{tab:knecht} are based on a reduced dataset with the last time step removed in order to permit out-of-sample prediction. To get as close as possible to the SAOM's rate-of-change function and mini-steps, a dyadic stability term (Equation~\ref{stability}) is added to the TERGM, which accounts for the inertia of both edges and non-edges in the data-generating process. While a SAOM models the change between time steps $t - 1$ and $t$, a TERGM models the state of the network at $t$, and therefore the change between $t - 1$ and $t$ can be introduced into the model by conditioning on the edges and non-edges of the network at $t - 1$ via the dyadic stability term. The SAOM is estimated by generalized method of moments (GMM) as is standard in the SAOM literature and the default estimation option for the software. The TERGM is estimated by MCMC-MLE. As the same information (the previous and current networks) and the same number of parameters (the parameters for the model terms and the rate parameter versus the stability term) enter both models, there is no reason to expect a priori that either of the two models should predict the last time step more successfully than the other model with an identical set of model terms, unless the assumptions the respective model makes do not capture the true data-generating process.

\begin{table}
\caption{Re-analysis of ``Model 0'' from \citet{snijders2010introduction} using the SAOM and TERGM, based on the first three time steps of the Knecht data.\medskip}
\label{tab:knecht}
\begin{center}
\begin{tabular}{l D{)}{)}{11)3} D{)}{)}{12)3} }
\toprule
 & \multicolumn{1}{c}{SAOM} & \multicolumn{1}{c}{TERGM} \\
\midrule
Density/edges             & -2.01 \; (0.60)^{***} & -10.45 \; (0.95)^{***} \\
Reciprocity               & 1.54 \; (0.29)^{***}  & 2.43 \; (0.38)^{***}   \\
Transitive triplets       & 0.34 \; (0.06)^{***}  & 0.06 \; (0.05)         \\
Cyclic triplets           & -0.27 \; (0.10)^{**}  & -0.49 \; (0.13)^{***}  \\
Transitive ties           & 0.60 \; (0.27)^{*}    & 0.14 \; (0.21)         \\
Indegree of alter (sqrt)  & 0.08 \; (0.18)        & 1.44 \; (0.24)^{***}   \\
Outdegree of alter (sqrt) & -0.56 \; (0.23)^{*}   & -0.02 \; (0.13)        \\
Outdegree of ego (sqrt)   & 0.04 \; (0.11)        & 1.81 \; (0.21)^{***}   \\
Same primary school       & 0.45 \; (0.17)^{**}   & 0.65 \; (0.26)^{*}     \\
Male (alter)              & -0.03 \; (0.17)       & 0.29 \; (0.24)         \\
Male (ego)                & 0.30 \; (0.16)        & 0.61 \; (0.23)^{**}    \\
Male (match)              & 0.69 \; (0.16)^{***}  & 1.81 \; (0.27)^{***}   \\
Rate parameter period 1   & 9.27 \; (1.68)^{***}  &                        \\
Rate parameter period 2   & 9.36 \; (1.66)^{***}  &                        \\
Dyadic stability          &                       & 1.09 \; (0.13)^{***}   \\
\bottomrule
\multicolumn{3}{l}{\scriptsize{$^{***}p<0.001$, $^{**}p<0.01$, $^*p<0.05$}}
\end{tabular}
\end{center}
\end{table}

The results differ substantially between the two models. Neither model shows signs of degeneracy (see Appendix~\ref{mcmcdiag} for degeneracy diagnostics). The coefficients are scaled differently and so should not be compared directly, but in several cases even the direction and significance of estimates is reversed. In particular, transitive triplets, transitive edges, indegree of alter, and outdegree of ego yield substantively different conclusions. If these effects were of substantive interest to the researcher, this divergence would be alarming. Furthermore, the estimates of both models are roughly comparable with the estimates gathered from identical models applied to all four time points. This indicates that the data-generating process does not change between the first three waves and the last wave, which is used for out-of-sample prediction.

Although we could try to tweak the two models and deviate from the model suggested by the original authors in order to improve model fit altogether, this would make our analysis vulnerable to criticism because doing so may favor one of the models over the other. Therefore we try to remain as impartial as possible by re-using the theoretical specification proposed by \citet{snijders2010introduction}.

\begin{figure}[t]
\begin{center}
\includegraphics[width=1.0\textwidth]{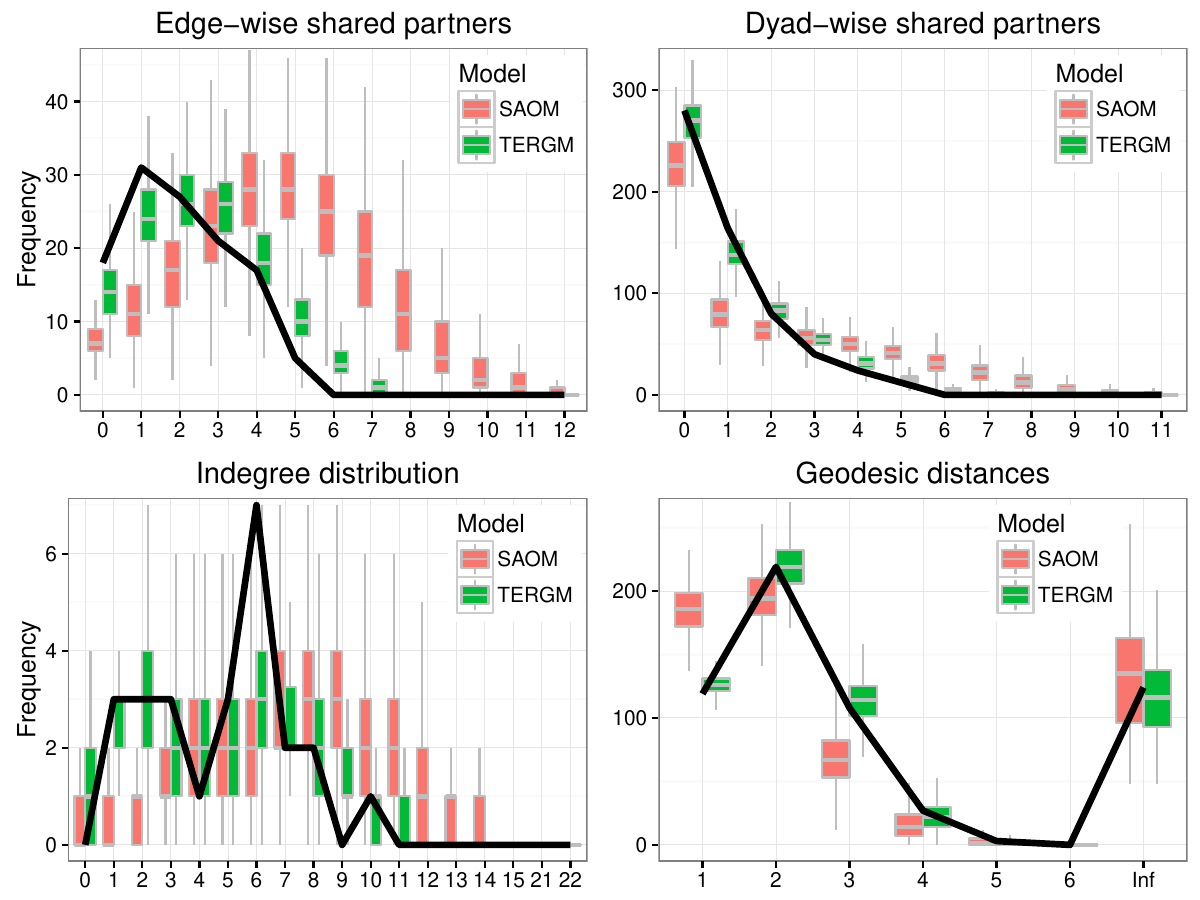}
\end{center}
\caption{Out-of-sample fit according to auxiliary network statistics. Green (red) boxplots indicate the distributions of network statistics for simulations based on the estimated TERGM (SAOM) model. The TERGM fits better on average in this empirical application.}
\label{fig:knecht-boxplot}
\end{figure}

\begin{figure}[t]
\begin{center}
\includegraphics[width=1.0\textwidth]{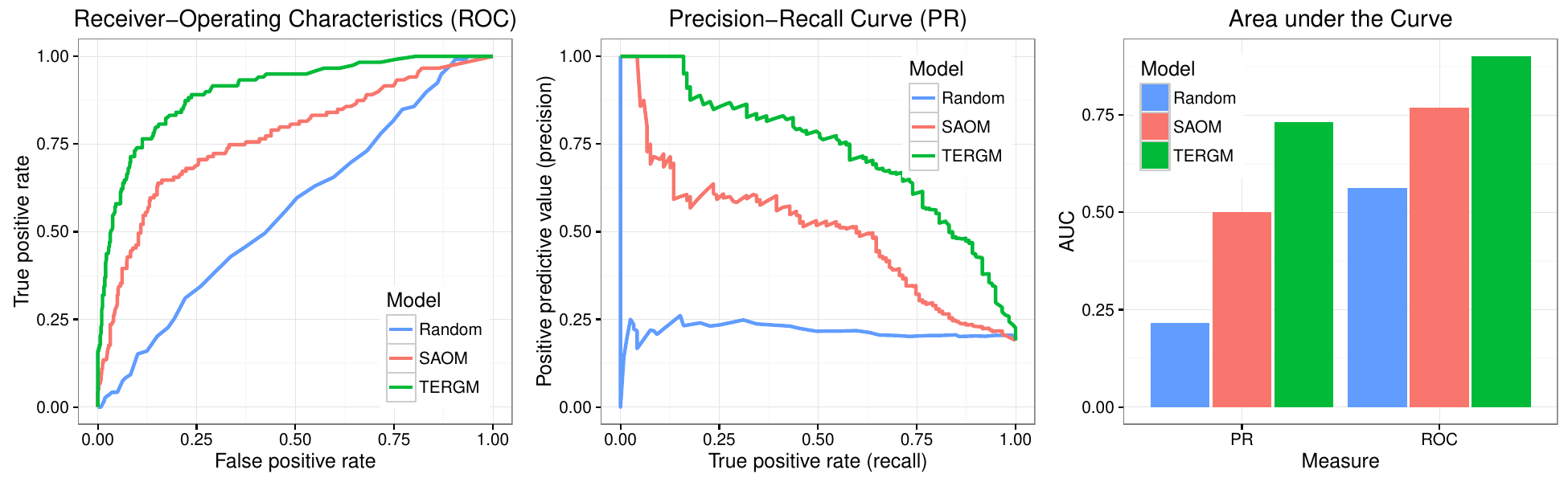}
\end{center}
\caption{Out-of-sample fit according to the receiver operating characteristic (ROC) curve (left panel) and precision--recall (PR) curve (center panel), and aggregate comparison of the area under the curve (AUC) for all models and curves.}
\label{fig:knecht-rocpr}
\end{figure}

We use out-of-sample prediction of time step~4 to assess which model fits the data best.
To do so, we estimate the models without information on the network at the last time step (as reported in Table~\ref{tab:knecht}), simulate 100 new networks in lieu of the last time step and based on the model, and compare these simulations to the observed network at the fourth time step. Given that we have a theoretically informed model as specified by \citet{snijders2010introduction}, comparable right-hand-side specifications, and real data that correspond well in principal to the SAOM's data generating process, the model that produces better out-of-sample predictive fit is the model doing a better job of capturing the data generating process that exists in nature. Figure~\ref{fig:knecht-boxplot} presents boxplots of endogenous network statistics (``auxiliary statistics'') in the simulated networks (as shown by the colored boxplots) versus the observed network (indicated by a black line).
This approach focuses on the structure of the network irrespective of the ordering of the actual vertices.
Very good fit occurs when the observed data intersect the medians of the predicted data.
Figure~\ref{fig:knecht-rocpr} shows the ROC and PR curves as well as the area under these curves as complementary measures of the goodness of fit.
These measures take into account the ordering of the actual vertices and assess classification performance of the models, that is, the number of dyads predicted correctly by the simulations.

The comparison of the boxplots across the two models reveals that the TERGM fits substantially better than the SAOM in this case. With all four auxiliary statistics, the simulated networks capture the observed network much more closely. Overall, there is some small room for improvement even in the TERGM, but the fit seems adequate for an out-of-sample prediction.

The ROC and PR comparison also confirms that the TERGM has a better out-of-sample model fit, by a large margin.
The TERGM outperforms the SAOM presented here significantly regarding the state of the specific dyads, no matter whether non-edges are part of the prediction task (as in the ROC curve) or not (as in the PR curve).

The difference in model fit is presumably at least partially achieved by the dyadic stability term, which conditions on the previous state of the network. The SAOM fails to add this simple explanation to the assumed DGP by design of its updating process. At the same time, this shows that the transitivity effects of the SAOM as shown in Table~\ref{tab:knecht} are spurious when controlling for the previous state of the network.

Three lessons can be learned from this replication exercise, though the lessons of an individual replication do not necessarily apply to other cases.
First, even in a social context around which the SAOM was originally designed, the TERGM can sometimes outperform the SAOM with regard to predictive performance.
This is particularly noteworthy because both models were parameterized as similarly as possible.
The only differences are the core assumptions inherent in the two models: the rate function and the mini-step updating process of the SAOM versus the dyadic stability term added to the TERGM.
Just like the updating process in the SAOM, the dyadic stability term in the TERGM is a very specific way of modeling time dependencies.
Yet the TERGM is considerably more flexible as to what kinds of temporal dependencies can be added should the data-generating process turn out to be of a different kind (e.\,g., edge innovation or loss, autoregression, delayed reciprocity etc.).
Here, merely conditioning on the previous time step explicitly yields a better prediction of the successive time step than Markovian mini-step updating from the previous time step onwards.
However, it is not clear whether this is the only or even the decisive mechanism leading to different model fit as the SAOM is built around a range of specific assumptions any of which could have led the updating process astray.

Second, even though we had no reason to cast doubt on the reasonableness of the SAOM updating process in the case of friendship networks a priori, it turned out that the estimates and simulations produced by the TERGM are more closely aligned with the oberved network. This is somewhat surprising and leads us to conclude that a priori model choice cannot always be informed by theory alone, at least given our current knowledge as to what is causing the limitations of the SAOM in this specific case. Candidates were enumerated above in the theoretical section. This underlines, once more, the need for more focused simulation experiments on each of the differences between the models separately.

And third, since theory alone cannot be evoked at the present time to choose one model over the other, we suggest the use of out-of-sample prediction in any given application to inform model choice for the time being. Our companion software 
([blinded])
makes such comparisons trivial. If many such comparisons are reported, this may soon lead to a better collective understanding of the contexts in which the TERGM is indeed to be preferred over the SAOM and vice versa and therefore provide complementary evidence to the simulation experiments suggested above.

\section{Conclusion}
We have shown that the TERGM and SAOM share a similar mathematical core, but differ with respect to several important assumptions about how the network evolves between observations. More specifically, the SAOM posits a particular process by which an observed network at $t-1$ transitions into the observed network at $t$. This process involves two stochastic sub-processes in which (a) vertices are selected for a chance to update their edge profiles and (b) the choice to change (make a new or dissolve an existing) an edge or not is made. The TERGM, conversely, is less specific. Though the TERGM is potentially consistent with a broader set of dynamic network processes, the SAOM could be though more appropriate in cases where its updating process is consistent with theory.

We conducted two comparative analyses of the TERGM and SAOM. In the first, we simulated data from a simple, entirely endogenous data-generating process. This was an instructive exercise because the ground truth was known, and we could judge clearly which models recovered coefficients closest to the actual data-generating process. Here, we found that both models performed similarly with regard to endogenous network properties but differently and in line with their respective assumptions with regard to tie prediction. Second, we considered a real-data example that maximally conforms to the updating process and application area for which the SAOM was designed: a friendship network of students in a Dutch school. In the real-data application, we replicated a SAOM that is commonly used in \texttt{RSiena} documentation and workshops. We found that, when considering out-of-sample predictive performance, the TERGM outperformed the SAOM by a substantial margin, both with regard to endogenous structures and the location of edges in the network.

From a theoretical perspective, a tentative conclusion we might reach, given the core similarity of the two models but the detailed updating process proffered by the SAOM, is that the SAOM should perform well when its core assumptions are met but the TERGM may perform better when they are not. This is supported by the simulation study where the SAOM and the TERGM could recover their respective underlying DGPs best. This conclusion, though, is in tension with our real-data example: indeed, even on the type of data for which the SAOM was designed and which the developers of the SAOM use as an expository case, the TERGM out-performed the SAOM. Yet it is not the case that the TERGM \emph{always} outperforms the SAOM: even with the TERGM DGP, a small fraction of the simulations were apparently more readily explained by the updating process of the SAOM (while the majority of these simulations could be better predicted by the TERGM). Our conclusion thus must be that the need of the SAOM to have its updating assumptions met with a high degree of precision is essential in order for that specific model to outperform the more general TERGM.

This discussion illustrates that it is usually not possible to choose the SAOM or the TERGM over their respective alternative purely on theoretical grounds. This may be true even when the researcher has certain a priori expectations with regard to the updating process and the degree to which the assumptions of each model are or are not met, as illustrated in the empirical replication exercise. Because theoretical conformity to the data-generating process of the SAOM did not yield predictable results, we must be careful not to put too much stock into the \emph{a priori} selection of a model. Rather, as it is straightforward to contrast the out-of-sample (or in-sample) predictive performance of the two models---the 
\texttt{xergm}
package has easy-to-use functions for making such comparisons---, the prudent researcher may be advised to do so.
Future applications of statistical network models for discretely observed dynamic networks should therefore by default take into account their respective alternative(s) and produce theoretical knowledge through comparison of out-of-sample predictive fit where possible, or at least report the results of both models (provided that both can be estimated).

\enlargethispage{1cm}

We have employed a range of methods in comparing predictive fit: side-by-side comparison of endogenous model fit through boxplots as in Figure~\ref{fig:knecht-boxplot}; edge prediction curves as in Figure~\ref{fig:knecht-rocpr}; and a direct comparison through subtracting deviations between the TERGM and the observed value from deviations between the SAOM and the observed value on the respective statistic (Equation~\ref{eq:diff} and Figure~\ref{fig:simulation-endogenous}). Applied researchers can use these methods for model comparison. Future research should explore more formal significance test procedures for making such comparisons, for example through a one-tailed $t$-test of the differences in Equation~\ref{eq:diff} from zero or through more elaborate means.

Finally, we have only touched the surface of this topic by contrasting carefully selected extreme cases. Much more work needs to be done in order to carve out which of the specific parts of the models lead to the differences we could observe. Our theoretical section offers a number of suggestions on factors that could make a difference. Future research will need to evaluate these factors as impartially as possible using simulation experiments and possibly carefully selected case studies.

\bibliographystyle{apsr}
\bibliography{literature}

\newpage

\appendix
\section{Online Appendix}

\subsection{Alternative Memory Terms for the TERGM}
As memory terms in TERGMs have not received extensive treatment outside of
\citet{Leifeld:2018},
the reader may wish to consider several example memory terms beyond what we discussed in the main text. The memory terms considered below are based on those in the extensive discussion from
\citet{Leifeld:2018}
and are by no means exhaustive, as a memory term can include arbitrary functions of time.

Several intuitive and convenient memory terms are as follows:

\begin{enumerate}
\item \emph{Positive autoregression (lagged outcome network):} $\textrm{h}_a = \sum N_{ij}^{t}N_{ij}^{t-1}.$ \\ This memory term adds value to the statistic any time an edge persists from one period to the next. Note that it can be modified by adding dependencies on networks further removed than $K=1$. Note, too, that, as formulated above, this is equivalent to adding a lagged network as a covariate in the TERGM.
\item \emph{Dyadic stability:}  $\textrm{h}_s = \sum_{ij} N_{ij}^{t} N_{ij}^{t-1} + (1- N_{ij}^{t}) (1-N_{ij}^{t-1})$\\
Dyadic stability is a straightforward extension of positive autocorrelation, but accounting for positive autocorrelation in nonexistent edges as well as existing ones. This simple statistic was also discussed in the main text.
\item \emph{Edge innovation/loss:}  $\textrm{h}_e = \sum_{ij} N_{ij}^{t} (1 - N_{ij}^{t-1})$ (innovation) and $\textrm{h}_l= \sum_{ij} (1- N_{ij}^{t})N_{ij}^{t-1}$ (loss).\\
These memory terms focus on the formation of new edges (innovation) or the dissolution of existing edges (loss) between time periods.

\item \emph{Delayed reciprocity (ego delays):} $\textrm{h}_a = \sum N_{ij}^{t}N_{ji}^{t-1}.$ \\
This statistic skirts the edge between a memory term and an endogenous dependency. It is a reciprocity statistic in which the ego reciprocates only after the alter has formed an edge to it. An alter-delays version may be created by swapping the order of $i$ and $j$ in the subscripts.
\end{enumerate}

\subsection{MCMC Diagnostics for the Empirical Model} \label{mcmcdiag}
\begin{figure}[tp]
\begin{center}
\includegraphics[width=1.0\textwidth]{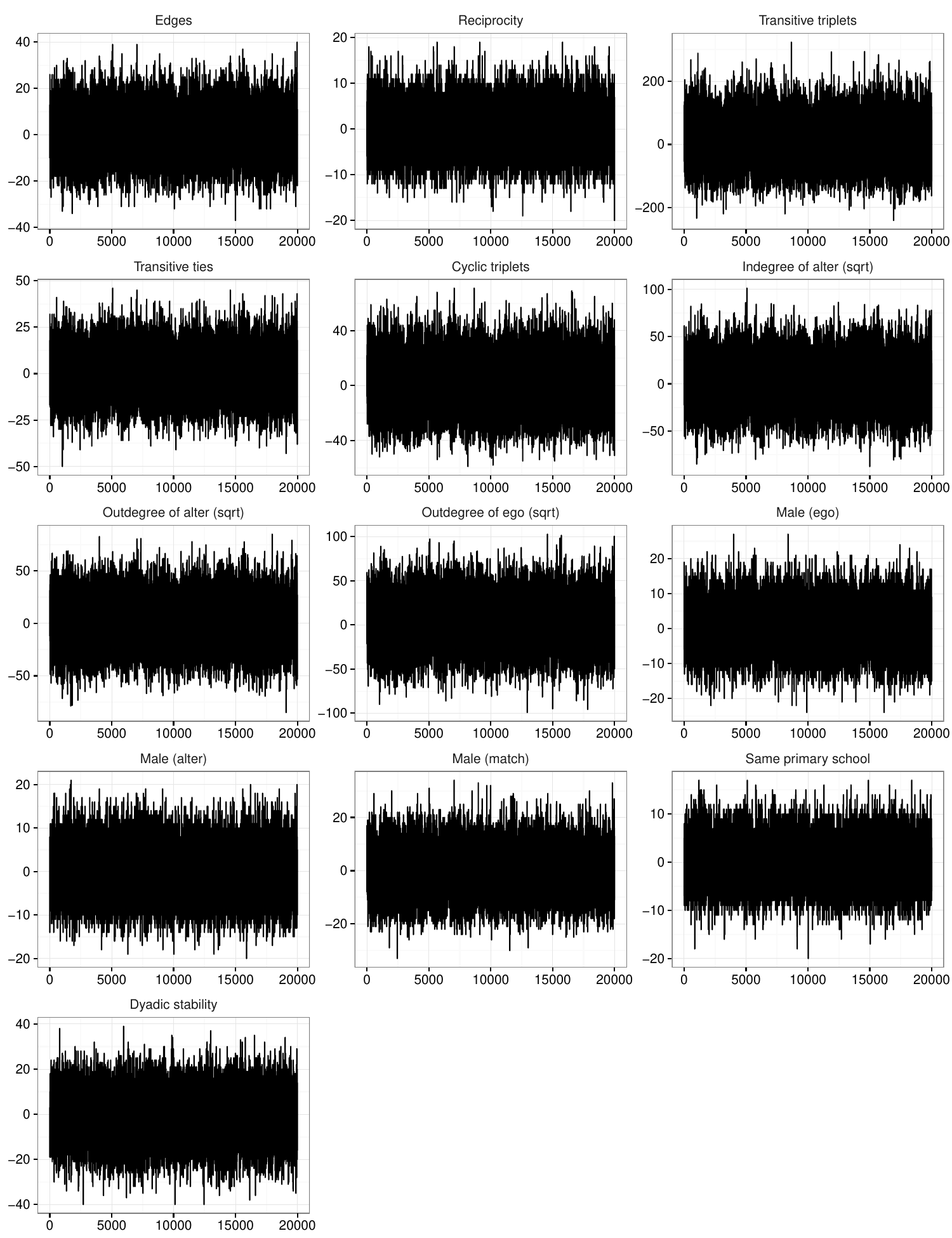}
\end{center}
\caption{MCMC trace plots for the empirical TERGM. The MCMC traces show that the empirically estimated TERGM is not degenerate as all statistics are in a stationary distribution over the duration of the 20,000 iterations.}
\label{fig:knecht-mcmctrace}
\end{figure}

\begin{figure}[tp]
\begin{center}
\includegraphics[width=1.0\textwidth]{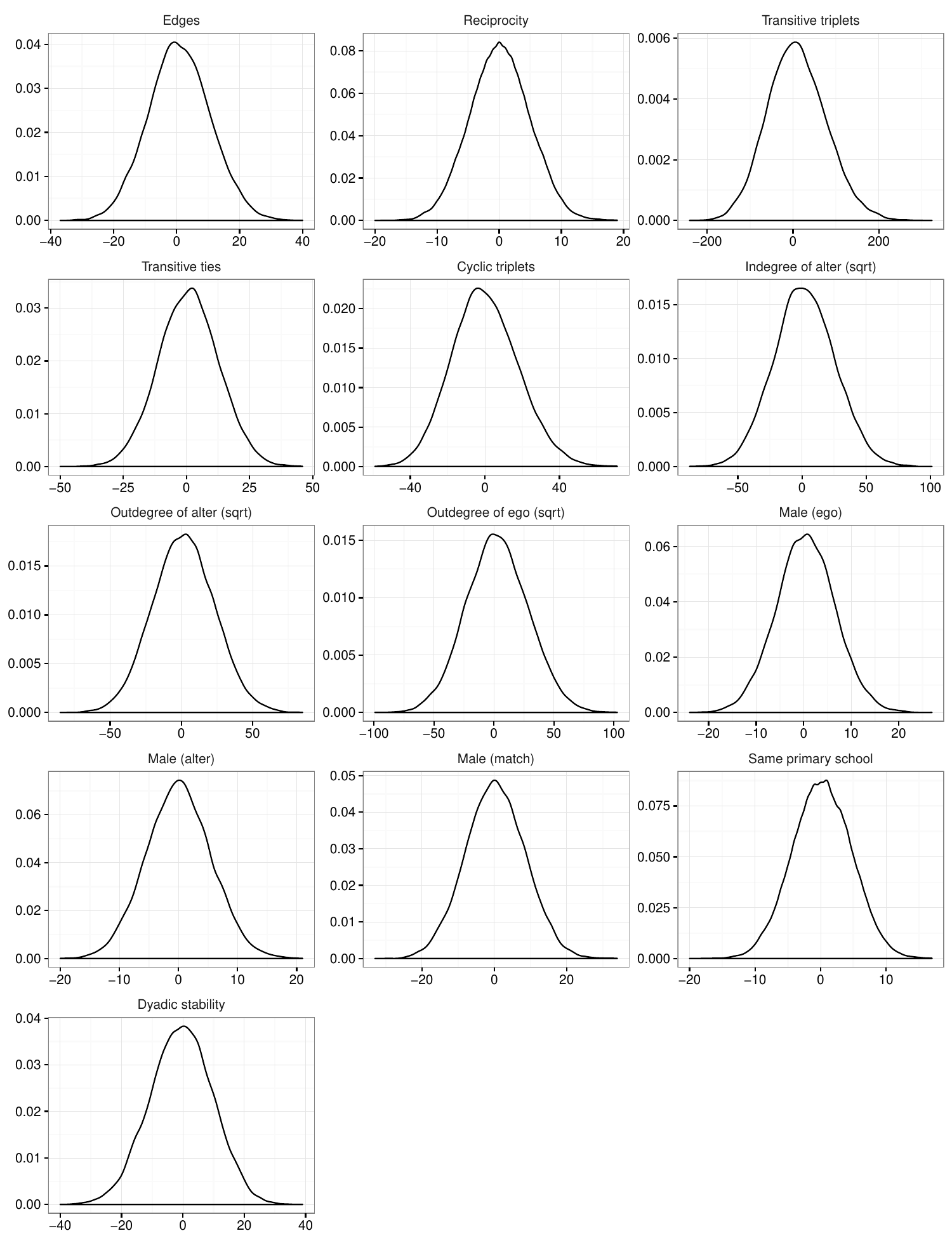}
\end{center}
\caption{MCMC density plots for the empirical TERGM. The density curves show the distribution of deviations along the time axis in Figure~\ref{fig:knecht-mcmctrace} and indicate that the empirically estimated TERGM is not degenerate as all deviations are approximately normally distributed.}
\label{fig:knecht-mcmcdensity}
\end{figure}

Figures~\ref{fig:knecht-mcmctrace} and~\ref{fig:knecht-mcmcdensity} show degeneracy checks for the empirical TERGM and indicate that the model is not degenerate.

\end{document}